\documentclass[twocolumn]{aastex701}

\begin{document}
	
\title{\textbf{\large{
Comparing the Near-infrared Spectral Energy Distributions from Different Stellar Population Synthesis Models with SPHEREx Observations
}}}

\author[0000-0003-3301-759X]{Jeong Hwan Lee}
\affil{Research Institute of Basic Sciences, Seoul National University, Seoul 08826, Republic of Korea}
\affil{Department of Physics and Astronomy, Seoul National University, 1 Gwanak-ro, Gwanak-gu, Seoul 08826, Republic of Korea}
\email{joungh93@gmail.com}
 
\author[0000-0002-3560-0781]{Minjin Kim}
\affil{Department of Astronomy, Yonsei University, 50 Yonsei-ro, Seoul 03722, Republic of Korea}
\email{mkim.astro@gmail.com}

\author[0000-0002-2770-808X]{Woong-Seob~Jeong}
\affil{Korea Astronomy and Space Science Institute, Daedeok-daero 776, Yuseong-gu, Daejeon 34055, Republic of Korea}
\email{}

\author[0000-0003-3078-2763]{Yujin~Yang}
\affil{Korea Astronomy and Space Science Institute, Daedeok-daero 776, Yuseong-gu, Daejeon 34055, Republic of Korea}
\email{}

\author[0000-0001-5382-6138]{Daniel~C.~Masters}
\affil{IPAC, California Institute of Technology, MC 100-22, 1200 East California Boulevard, Pasadena, CA 91125, USA}
\email{}

\author[0000-0002-5037-951X]{Kyuseok~Oh}
\affil{Korea Astronomy and Space Science Institute, Daedeok-daero 776, Yuseong-gu, Daejeon 34055, Republic of Korea}
\email{}

\author[0000-0003-1954-5046]{Bomee~Lee}
\affil{Korea Astronomy and Space Science Institute, Daedeok-daero 776, Yuseong-gu, Daejeon 34055, Republic of Korea}
\email{}

\author[0000-0001-7432-2932]{Olivier~Dor\'{e}}
\affil{Jet Propulsion Laboratory, California Institute of Technology, 4800 Oak Grove Drive, Pasadena, CA 91109, USA}
\affil{Department of Physics, California Institute of Technology, 1200 East California Boulevard, Pasadena, CA 91125, USA}
\email{}

\author[0009-0009-1219-5128]{Zhaoyu~Huai}
\affil{Department of Physics, California Institute of Technology, 1200 East California Boulevard, Pasadena, CA 91125, USA}
\email{}

\author[0000-0002-5437-0504]{Yun-Ting~Cheng}
\affil{Jet Propulsion Laboratory, California Institute of Technology, 4800 Oak Grove Drive, Pasadena, CA 91109, USA}
\affil{Department of Physics, California Institute of Technology, 1200 East California Boulevard, Pasadena, CA 91125, USA}
\email{}

\author[0000-0002-9330-8738]{Richard~M.~Feder}
\affil{University of California at Berkeley, Berkeley, CA 94720, USA}
\email{}

\author[0000-0001-8253-1451]{Michael~Zemcov}
\affil{Jet Propulsion Laboratory, California Institute of Technology, 4800 Oak Grove Drive, Pasadena, CA 91109, USA}
\affil{School of Physics and Astronomy, Rochester Institute of Technology, 1 Lomb Memorial Drive, Rochester, NY 14623, USA}
\email{}

\author[0000-0003-1647-3286]{Yongjung~Kim}
\affil{School of Liberal Studies, Sejong University, 209 Neungdong-ro, Gwangjin-Gu, Seoul 05006, Republic of Korea}
\affil{Korea Astronomy and Space Science Institute, Daedeok-daero 776, Yuseong-gu, Daejeon 34055, Republic of Korea}
\email{}

\author[0000-0002-6925-4821]{Dohyeong~Kim}
\affil{Department of Earth Sciences, Pusan National University, Busan 46241, Republic of Korea}
\email{}

\author[0000-0002-8055-5465]{Jong-Hak~Woo}
\affil{Department of Physics and Astronomy, Seoul National University, 1 Gwanak-ro, Gwanak-gu, Seoul 08826, Republic of Korea}
\email{}

\author[0000-0002-9382-9832]{Andreas~L.~Faisst}
\affil{IPAC, California Institute of Technology, MC 100-22, 1200 East California Boulevard, Pasadena, CA 91125, USA}
\email{}

\author[0000-0001-5812-1903]{Howard~Hui}
\affil{Jet Propulsion Laboratory, California Institute of Technology, 4800 Oak Grove Drive, Pasadena, CA 91109, USA}
\affil{Department of Physics, California Institute of Technology, 1200 East California Boulevard, Pasadena, CA 91125, USA}
\email{}

\author[0000-0002-4650-8518]{Brendan~P.~Crill}
\affil{Jet Propulsion Laboratory, California Institute of Technology, 4800 Oak Grove Drive, Pasadena, CA 91109, USA}
\affil{Department of Physics, California Institute of Technology, 1200 East California Boulevard, Pasadena, CA 91125, USA}
\email{}

\author[0000-0001-9368-3186]{Chi~H.~Nguyen}
\affil{Department of Physics, California Institute of Technology, 1200 East California Boulevard, Pasadena, CA 91125, USA}
\email{}

\author[0000-0002-3892-0190]{Asantha~Cooray}
\affil{Department of Physics \& Astronomy, University of California Irvine, Irvine, CA 92697, USA}
\email{}


\correspondingauthor{Minjin Kim}
\email{mkim.astro@gmail.com}

\begin{abstract}

While stellar population synthesis (SPS) models have been widely used for spectral analysis in optical wavelengths, their characteristics remain uncertain in the near-infrared (NIR) due to a relative lack of observed NIR spectra.
The spectrophotometric data from SPHEREx are well-suited for investigating the performance of SPS models in the NIR, thanks to its wide wavelength coverage over $0.75-5.0~{\rm \mu m}$.
In this work, we compare the observed SPHEREx data of SDSS compact galaxies, including 2,726 non-emission-line galaxies and 1,163 emission-line galaxies, to the NIR SEDs predicted from the full spectrum fitting of SDSS optical spectra.
We use four different SPS models that extend into the NIR: E-MILES, Bruzual \& Charlot (BC03), Charlot \& Bruzual (CB19), and FSPS.
We find that all four models tend to overpredict the stellar continuum at $2.4-5~{\rm \mu m}$ by $0.1-0.3~{\rm mag}$.
This trend is particularly prominent for intermediate-age stellar populations ($\sim1-5~{\rm Gyr}$), suggesting a systematic bias in the NIR SED predictions of current SPS models.
For stellar populations older than $5~{\rm Gyr}$, E-MILES shows relatively smaller offsets at $3.8-5~{\rm \mu m}$ compared to other models.
Meanwhile, for emission-line galaxies, the SPS models underestimate the SED by up to $\sim0.5~{\rm mag}$ at longer wavelengths due to the contribution of non-stellar emission.
Overall, these results highlight the necessity of refining the NIR stellar spectral features in SPS models, such as emissions from thermally pulsating asymptotic giant branch stars or molecular absorptions from cool stars.

\end{abstract}

\section{Introduction}

Stellar population synthesis (SPS) models provide the fundamental framework for investigating the stellar population properties of galaxies.
SPS models are composed of libraries of simple stellar population (SSP) templates spanning a range of stellar ages and metallicities, which predict the spectral energy distributions (SEDs) of composite stellar populations \citep{con13}.
These models have been widely used in observational studies, including SED fitting with broadband photometric data \citep{wal11, pac12, boq19} and full-spectrum fitting of spectroscopic data \citep{cid05, kol08, wil17, cap23}.
These techniques have allowed us to derive physical properties such as stellar age, metallicity, star formation history (SFH), dust attenuation, and mass-to-light ratios.
Therefore, a detailed understanding of SPS models is essential for analyzing the stellar populations of galaxies.

The SSP templates within SPS models are constructed from several key components, including initial mass functions (IMFs), stellar evolutionary isochrones, stellar spectral libraries, and prescriptions of underlying stellar physics (see \citeauthor{con13} \citeyear{con13} for review).
Depending on the combinations of these ingredients, various SPS models have been developed for observational analysis: PEGASE \citep{fio97}, Bruzual \& Charlot 2003 (\citeauthor{bru03} \citeyear{bru03}; BC03 hereafter), Maraston 2005 \citep{mar05}, Starburst99 \citep{vaz05}, FSPS \citep{con09, con10a}, Maraston 2011 \citep{mar11}, E-MILES \citep{vaz16}, BPASS \citep{eld17}, Charlot \& Bruzual 2019 (\citeauthor{pla19} \citeyear{pla19}; CB19 hereafter), and MaStar \citep{yan19, mar20}.
The diverse model ingredients and procedures used in SPS models can introduce systematic differences in the inferred SEDs and stellar population properties.
Previous studies have reported variations in optical and near-infrared (NIR) colors among BC03, FSPS, and Maraston 2005 \citep{con10b} and estimates for ages and metallicities of galaxies depending on the adopted SSP library \citep{ge19}.
The impacts of these differences can affect a wide range of astrophysical fields; for example, the choice of SPS model can lead to systematic uncertainties in the measurement of the Hubble parameter, $H(z)$, and the predicted age--$D_{n}4000$ relation \citep{mor20}.

In particular, systematic differences in the predicted SEDs among SPS models appear to be more pronounced in the NIR than in the optical wavelengths.
\citet{man12} compared the magnitudes and colors predicted from five SPS model sets and revealed that the scatter in predicted magnitudes is relatively small in the SDSS optical filters ($\sigma\sim0.1~{\rm mag}$), but increases dramatically to $\sigma\sim0.7~{\rm mag}$ in the NIR $K_{s}$ band.
Similarly, \citet{bal18} explored how the predicted SFHs of 12 early-type galaxies depend on the choice of SPS models, using the full-spectrum fitting of optical and NIR spectra from the ATLAS${\rm ^{3D}}$ survey and Gemini GNIRS observations.
They found that the SFHs derived from different SPS models are in agreement when using optical spectra, whereas those derived using NIR spectra vary significantly with the adopted models.
These findings highlight the substantial differences in the predicted NIR SEDs among SPS models, making it crucial to address the uncertainties in modeling the NIR-bright stellar populations.

One of the main reasons for these NIR discrepancies is the treatment of emission from thermally pulsating asymptotic giant branch (TP-AGB) stars.
These intermediate-age ($\sim0.2-2~{\rm Gyr}$) stars represent a late evolutionary stage of cool ($T_{e} \lesssim 4000~{\rm K}$) giant stars with masses ranging from $\sim1-8~{M_{\odot}}$, where stars are undergoing recurrent thermonuclear flashes in the H- and He-burning shells (see \citeauthor{her05} \citeyear{her05} for review).
It has been challenging to model the emission from TP-AGB stars because of their small observational sample sizes and complicated physical processes, such as the third dredge-up and strong mass loss by stellar winds \citep{con13, kar14, hof18}.
The emission from TP-AGB stars can contribute up to $\sim80\%$ of the NIR flux of intermediate-age stellar populations \citep{mar98}, making them an important source of uncertainty in the predicted NIR SEDs of SPS models.
For instance, the Maraston 2005 model, which includes a strong TP-AGB contribution, predicts significantly brighter fluxes at $\sim 1-2.5~{\rm \mu m}$ for intermediate-age stellar populations compared to other SPS models such as BC03, PEGASE, and Starburst99, yielding different SED fitting results from local globular clusters \citep{mar05} to high-redshift galaxies \citep{mar06}.
In contrast, several studies have suggested that BC03 and FSPS models show a better match for the optical and NIR SEDs of observed post-starburst galaxies than the Maraston 2005 model \citep{con10b, kri10, zib13}.

Another factor contributing to the NIR discrepancy is the spectral resolution and wavelength coverage of the adopted stellar spectral libraries in SPS models.
\citet{bal18} showed that the SFHs derived from optical and NIR spectra become more consistent when high-resolution NIR spectral libraries ($R = \lambda / \Delta \lambda \sim2000$), such as IRTF \citep{cus05, ray09}, C3K \citep{con12, con18}, are used instead of lower-resolution libraries ($R\sim200-500$) such as BaSeL \citep{wes02} and Pickles \citep{pic98}.
In addition, SPS models exhibit different features of CO absorption bands emitted from late-type stars, including the first overtone band at $2.3-2.5~{\rm \mu m}$ and the fundamental band at $4.2-4.5~{\rm \mu m}$.
For example, the E-MILES model shows prominent fundamental CO band absorption at $4.2-4.5~{\rm \mu m}$, resulting in substantially bluer $[3.6]-[4.5]$ NIR colors compared to other SPS models \citep{LeeJH25}.
This difference arises because the E-MILES model extends its IRTF spectral library to the $2.5-5.0~{\rm \mu m}$ range using the high-resolution Phoenix library \citep{all12, roc15}, whereas other SPS models mostly adopted the lower-resolution BaSeL library at these wavelengths.
Taken together, the NIR differences among widely used SPS models emphasize the need for a comprehensive test using large galaxy samples to properly understand stellar population properties in the NIR.

The Spectro-Photometer for the History of the Universe, Epoch of Reionization, and Ices Explorer (SPHEREx) is a space telescope featuring a $20\,\mathrm{cm}$ entrance pupil equipped with three freeform aluminum mirrors to map the entire sky in the NIR wavelength regime.
It provides the first all-sky spectrophotometric datasets covering wavelengths from $0.75~{\rm \mu m}$ to $5.0~{\rm \mu m}$, offering a great opportunity to perform the comparative analysis of NIR-covering SPS models for large galaxy samples \citep{boc26}.
Using a linear variable filter (LVF) with a wide field of view ($3.5\times11.3~{\rm deg^{2}}$), the SPHEREx mission will survey the entire sky four times for two years with a dense NIR wavelength coverage through 102 spectral channels grouped into six wavelength bands.
The SPHEREx bands are defined as Band 1 ($0.75-1.11~{\rm \mu m}$), Band 2 ($1.11-1.64~{\rm \mu m}$), Band 3 ($1.64-2.42~{\rm \mu m}$), Band 4 ($2.42-3.82~{\rm \mu m}$), Band 5 ($3.82-4.42~{\rm \mu m}$), and Band 6 ($4.42-5.00~{\rm \mu m}$).
Bands 1--4 have relatively low spectral resolution ($R\sim35-40$) but high sensitivity ($\sim22~{\rm mag}$ at $5\sigma$ for the all-sky survey), while Bands 5 and 6 provide higher spectral resolution ($R\sim110-130$) with $\sim1-2~{\rm mag}$ shallower sensitivity than the shorter-wavelength bands \citep{boc26, hui26}.

In extragalactic fields, SPHEREx data enable systematic investigations of stellar mass maps \citep{LeeJH25}, photometric redshifts \citep{sti16, fed24, bae26}, polycyclic aromatic hydrocarbon (PAH) emissions at $3.3~{\rm \mu m}$ \citep{zha25}, and high-redshift quasars \citep{dav26}.
\citet{LeeJH25} has already performed a pilot study on how resolved stellar mass estimates of nearby late-type galaxies depend on the adopted NIR-covering SPS models.
By analyzing PHANGS-MUSE integral field spectroscopic data, they found that the BC03 and CB19 models underpredict the NIR stellar mass-to-light ratios by $0.2-0.3~{\rm dex}$ compared to the E-MILES and FSPS models.
Based on these results, \citet{LeeJH25} suggested different strategies for deriving resolved stellar masses from SPHEREx mock data depending on the choice of SPS models.

In this study, we use the SPHEREx Quick Release 2 (QR2) data to compare the NIR SEDs predicted by different NIR-covering SPS models with SPHEREx observations.
We consider four different NIR-covering SPS models, E-MILES, BC03, CB19, and FSPS, which are the same model sets as adopted in \citet{LeeJH25}.
Our analysis focuses on $\sim3900$ compact galaxies spanning a wide range of stellar ages and NIR colors.
These targets are intentionally selected to be unresolved at the SPHEREx pixel scale ($6\farcs15 \times 6\farcs15$), simplifying the SPHEREx spectrophotometric process and minimizing the systematic offsets in photometric data from SPHEREx and other instruments.
Following a similar approach to \citet{LeeJH25}, we perform full-spectrum fitting of the optical spectra of the galaxies to predict their best-fit NIR SEDs for each SPS model.
By comparing these predictions with SPHEREx observations, we quantify the systematic bias of NIR SEDs between the models and observations as functions of stellar age and NIR color of the galaxies.
Through these processes, this study evaluates the predictability of NIR-covering SPS models and discusses the implications of model selection on the stellar population analysis of galaxies in the NIR.

This paper is organized as follows.
In {\color{blue} \textbf{Section \ref{sec:data}}}, we describe our sample selection based on the Sloan Digital Sky Survey (SDSS) and explain how we obtain the NIR and mid-infrared (MIR) photometry from SPHEREx and the Wide-field Infrared Survey Explorer (WISE).
In {\color{blue} \textbf{Section \ref{sec:method}}}, we detail the methodology for full-spectrum fitting procedures and the statistical analysis used for the comparison.
Based on these processes, we present the results of our NIR SED comparisons in {\color{blue} \textbf{Section \ref{sec:result}}}.
In {\color{blue} \textbf{Section \ref{sec:discuss}}}, we discuss the relationship between the systematic biases and the stellar population properties of galaxies.
Finally, we summarize the main findings in {\color{blue} \textbf{Section \ref{sec:summary}}}.
Throughout the paper, SDSS magnitudes are on the AB system, while WISE magnitudes are given in the Vega system.
We adopt the $\Lambda$CDM cosmology with $H_{0}=70~{\rm km~s^{-1}~Mpc^{-1}}$, $\Omega_{M}=0.3$, and $\Omega_{\Lambda}=0.7$.

\section{Sample and Data}
\label{sec:data}

\subsection{Sample Selection}
\label{sec:sample}

In this study, we selected our galaxy sample using the SQL query of the SDSS Data Release 18\footnote[1]{\url{https://skyserver.sdss.org/dr18/SearchTools/sql}} \citep{alm23} to obtain both optical photometry and spectra.
We took a simple approach by focusing on compact galaxies to avoid complexities in SPHEREx photometry for extended sources \citep{ake25} and to minimize the discrepancies between SPHEREx and SDSS fluxes due to the aperture effects of SDSS fiber spectroscopy.
This selection was primarily motivated by instrumental considerations rather than by the intention to constrain any intrinsic physical properties of the galaxies.
Although the compact galaxies selected in this study might not be fully representative of general galaxy populations, they were not expected to have a significant selection bias on galaxy properties.

Using the \texttt{PhotoObjAll} and \texttt{SpecObjAll} tables, we first applied the following initial selection criteria: $i$-band fiber magnitude brighter than $20~{\rm mag}$ ($\texttt{fiberMag\_i}<20.0$), high spectral signal-to-noise ratio ($\texttt{snMedian\_r}>20$), spectroscopic classification as galaxies to exclude the effect from active galactic nuclei ($\texttt{class} = \texttt{`GALAXY'}$), and compact sizes with Petrosian radii smaller than $3\farcs1$ ($\texttt{petroRad\_[\textit{gri}]}<3.1~{\rm arcsec}$).
We also required $\texttt{z}>0.01$ and $\texttt{zWarning == 0}$ because the sources with $z < 0.01$ often tended to be misidentified with compact stellar regions or star clusters embedded in nearby galaxies.
To ensure high-quality photometry, we applied a clean photometry flag (\texttt{clean == 1}) and excluded objects with poor-quality photometric flags such as \texttt{SATURATED}, \texttt{EDGE}, \texttt{BLENDED}, \texttt{NODEBLEND}, and \texttt{COSMIC\_RAY}.
Additionally, we selected isolated sources with a minimum separation of $\texttt{distance} > 10~{\rm arcsec}$ from neighboring objects based on the \texttt{Neighbors} table to avoid any blending issues of SPHEREx \citep{dac24, hua26}.

\begin{figure}
\centering
\includegraphics[width=0.50\textwidth]{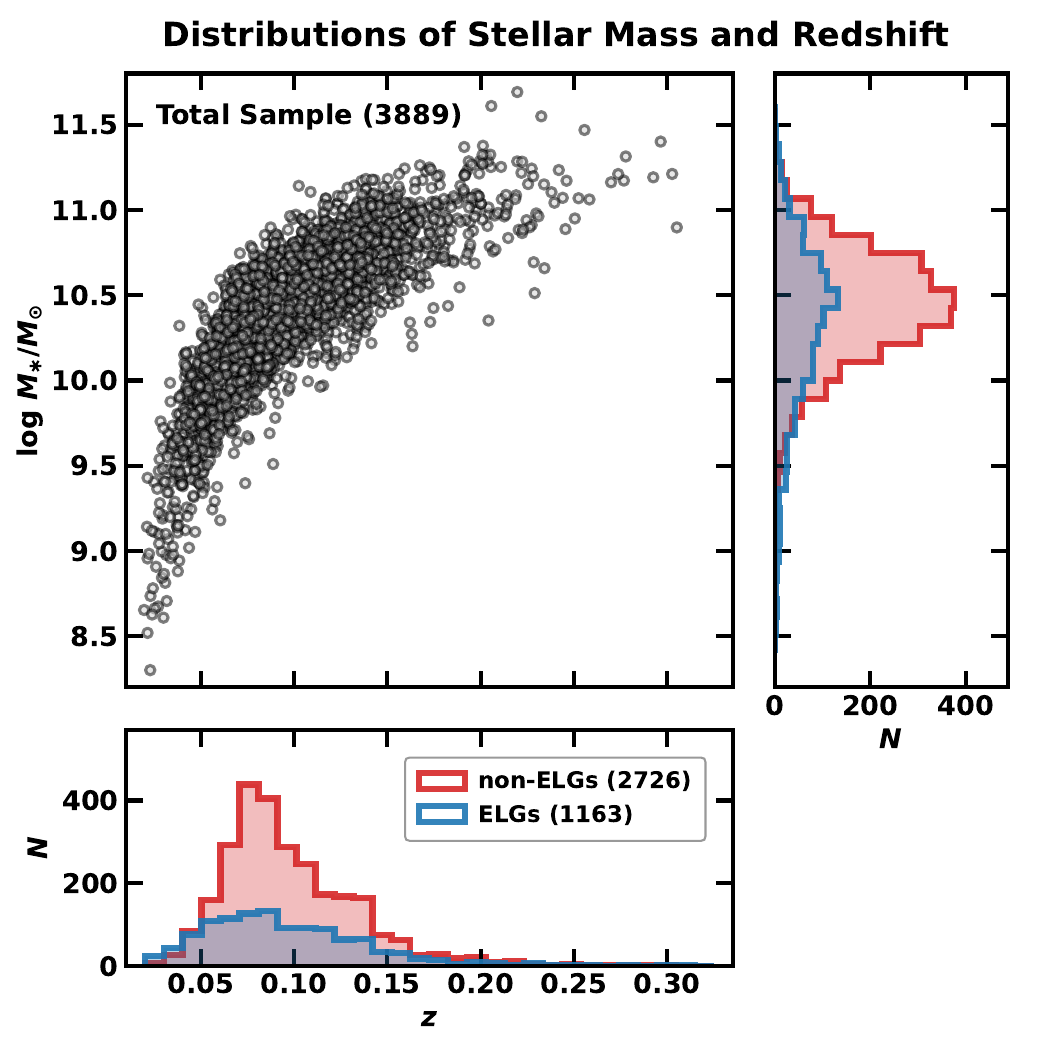}
\caption{
Distribution of stellar mass and redshift of the final galaxy sample used in our analysis.
In the main panel, the scatter plot presents the two-dimensional distribution of the total sample galaxies in the stellar mass--redshift plane.
The side panels show the corresponding histograms of redshift (bottom) and stellar mass (right) for the two subsamples of non-ELGs (red) and ELGs (blue).
\label{fig:dist1}}
\end{figure}

This initial selection process left 4,172 compact galaxy candidates with available SDSS optical photometry and spectra, spanning redshifts of $0.01 < z < 0.4$ and stellar masses of $8.5 < {\rm log}~M_{\ast}/M_{\odot} < 11.5$ (from the \texttt{stellarMassFSPSGranWideDust} table).
All these sample galaxies were observed with the SDSS-I/II spectrograph with a fiber diameter of 3 arcsec.

To correct for their foreground extinction, we adopted the reddening magnitudes from \citet{sch11} and the \citet{car89} extinction function, using the Python \texttt{dustmaps}\footnote[2]{\url{https://dustmaps.readthedocs.io/en/latest}} \citep{gre18} and \texttt{extinction}\footnote[3]{\url{https://extinction.readthedocs.io/en/latest}} \citep{bar16} packages.

\subsubsection{Galaxy Classification with Emission-line Features}

We further divided the galaxies into non-emission-line galaxies (non-ELGs) and emission-line galaxies (ELGs), based on the equivalent widths (EWs) of Balmer emission lines from the \texttt{galSpecLine} table: galaxies with ${\rm EW(H\alpha) < 3 \AA}$ and ${\rm EW(H\beta) < 2 \AA}$ were classified as non-ELGs, and the remaining galaxies with higher equivalent widths as ELGs.
In this process, we followed a generous classification for ELGs suggested in earlier works \citep{cid11, dua17, mar22}, considering the typical uncertainty level of the equivalent widths (${\rm \lesssim 1 \AA}$) and possible loss of the instrumental signals at either ${\rm H\alpha}$ or ${\rm H\beta}$ emission lines.
According to this classification, 2,866 galaxies were categorized as non-ELGs and 1,306 as ELGs among the 4,172 galaxies.
With this parent sample, we performed full-spectrum fitting to select galaxies with reliable kinematic and stellar population parameters (4,172 to 4,108; see {\color{blue} \textbf{Section \ref{sec:ppxf}}}) and cross-matched with their available SPHEREx dataset (4,108 to 4,066; see {\color{blue} \textbf{Section \ref{sec:spherex}}}).
These sample galaxies were further refined by excluding several galaxies with larger SPHEREx-to-SDSS flux offsets (4,066 to 3,889; see {\color{blue} \textbf{Section \ref{sec:offset}}}).
After these refinements, the final analysis sample consists of 3,889 compact galaxies with 2726 non-ELGs and 1163 ELGs.
The distributions of stellar mass and redshift for this sample are shown in {\color{blue} \textbf{Figure \ref{fig:dist1}}}.

This classification is intended to distinguish stellar and non-stellar contributions in the comparisons of NIR SEDs from SPHEREx and SPS models.
In this study, we constructed the best-fit SEDs from optical to NIR wavelengths for different SPS models using full spectrum fitting, as elaborated in {\color{blue} \textbf{Section \ref{sec:ppxf}}}.
It is noteworthy that these model-derived composite SEDs do not account for nebular or dust emission components, including only stellar emission from galaxies.
For instance, the PAH feature at $3.3~{\rm \mu m}$, emitted from interstellar dust around young stars, can significantly affect the observed NIR SEDs of star-forming galaxies but is not included in the model-derived SEDs.
Therefore, the discrepancies between the modeled SEDs and SPHEREx observations may depend not only on the choice of SPS models but also on the contribution of non-stellar emission from galaxies.
In this regard, our classification into non-ELGs and ELGs can allow us to particularly decompose the non-stellar contributions from ELGs, as ELGs are expected to exhibit more prominent nebular emission lines and dust continuum at $\lambda \gtrsim 2.5~{\rm \mu m}$.
As the main purpose of this work is to compare stellar SEDs from SPS models, we consider non-ELGs as the main analysis sample and ELGs as a supplementary control sample in the analysis.


\subsection{SPHEREx Dataset}
\label{sec:spherex}

We obtained SPHEREx data for the selected galaxies from the NASA/IPAC Infrared Science Archive (IRSA)\footnote[4]{\url{https://irsa.ipac.caltech.edu/applications/spherex}}.
The IRSA SPHEREx archive has provided SPHEREx QR2 data processed with the SPHEREx pipeline version 6.4 or later \citep{ake26}.
Using the IRSA Spectrophotometry Tool, we extracted NIR photometric measurements for our initial sample of the 4,172 compact galaxy candidates between 2026 April 14 and May 11.
Since our compact galaxy sample is expected to be unresolved at the SPHEREx pixel scale ($6\farcs15 \times 6\farcs15$), we adopted the ``point source'' profile for forced photometry, with the source positions fixed to the SDSS coordinates.
The background estimation region was set to a 15-pixel box, following the default configuration.

For robust measurements, we excluded objects with insufficient SPHEREx photometric coverage, requiring at least 10 valid data points in each SPHEREx band.
We only used SPHEREx data points with photometry flags of 0 (no flags) or 2097152 (bit 21; source) to exclude the problematic pixels affected by cosmic ray hits (bit 0), saturations (bit 1), dichroic mirrors between SPHEREx Bands 3 and 4 (bit 7), etc.
This left 4,066 galaxies among the initial SDSS sample.
We further refined this sample with the cross-calibration between SPHEREx and SDSS, as described in {\color{blue} \textbf{Section \ref{sec:offset}}}.

Individual spectra extracted from the IRSA Spectrophotometry Tool exhibit unique wavelength sampling because the wavelength information varies with the spatial position in the SPHEREx LVF planes.
To account for this, we resampled the SPHEREx data onto standardized bins of 102 wavelength channels to ensure a uniform comparison across the entire sample \citep{boc26, hui26}.
Using the standardized SPHEREx filter response curves, we interpolated the original SPHEREx photometric measurements onto this 102 unified wavelength grid for all galaxies.
This procedure facilitates a consistent quantitative comparison between modeled SEDs and SPHEREx observations.

\begin{figure*}
\centering
\includegraphics[width=0.85\textwidth]{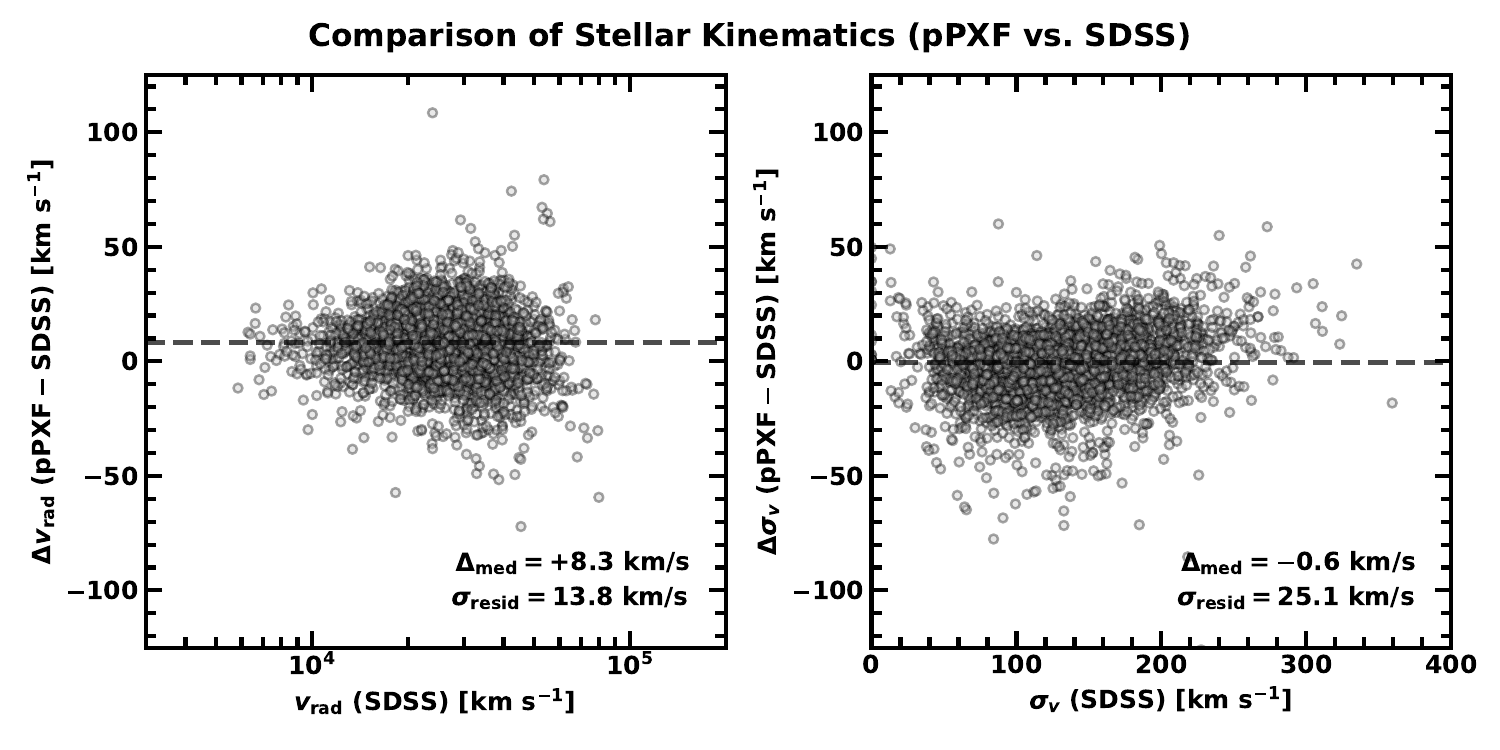}
\caption{
Comparisons of stellar kinematics of our sample galaxies, including radial velocity (left) and velocity dispersion (right), derived from pPXF in this work and values obtained from SDSS DR18.
Both panels show the SDSS measurements on the x-axis and the difference between the pPXF-derived kinematics and SDSS values on the y-axis.
The median values (black horizontal dashed lines) and standard deviations of the differences are noted in the bottom left side of the panels.
\label{fig:comp_kin}}
\end{figure*}

\subsection{WISE Photometry}
\label{sec:wise}

We retrieved additional WISE NIR/MIR photometric data from the IRSA archive\footnote[5]{\url{https://irsa.ipac.caltech.edu/applications/wise}}, based on the WISE All-Sky Release catalog \citep{wri10}.
We used these WISE data to determine broadband NIR and MIR colors of our sample galaxies and to compare their SPHEREx photometry with WISE measurements at overlapping wavelengths.
In particular, the color between the W1 ($3.4~{\rm \mu m}$) and W3 ($12~{\rm \mu m}$) bands serves as an effective proxy for star formation activity and therefore for robustly determining the mass-to-light ratio \citep{jar13, jar23, LeeJH25}.
This is because the W1 band primarily traces stellar continuum emission from late-type stars, whereas the W3 band is dominated by non-stellar emission from interstellar dust heated by young stars.
We therefore used the ${\rm W1-W3}$ color to investigate how non-stellar emissions contribute to the NIR SEDs, in particular for the ELG control sample.
For this analysis, we adopted WISE magnitudes derived from profile-fitting photometry \citep{cut12}.
Since our sample galaxies are also unresolved at the WISE point spread function (PSF) scale, with Full Widths at Half Maximum (FWHMs) of $\sim6''-7''$ in the W1, W2, and W3 bands, their integrated fluxes can be reliably measured by PSF profiles in each WISE band.

\begin{figure*}
\centering
\includegraphics[width=\textwidth]{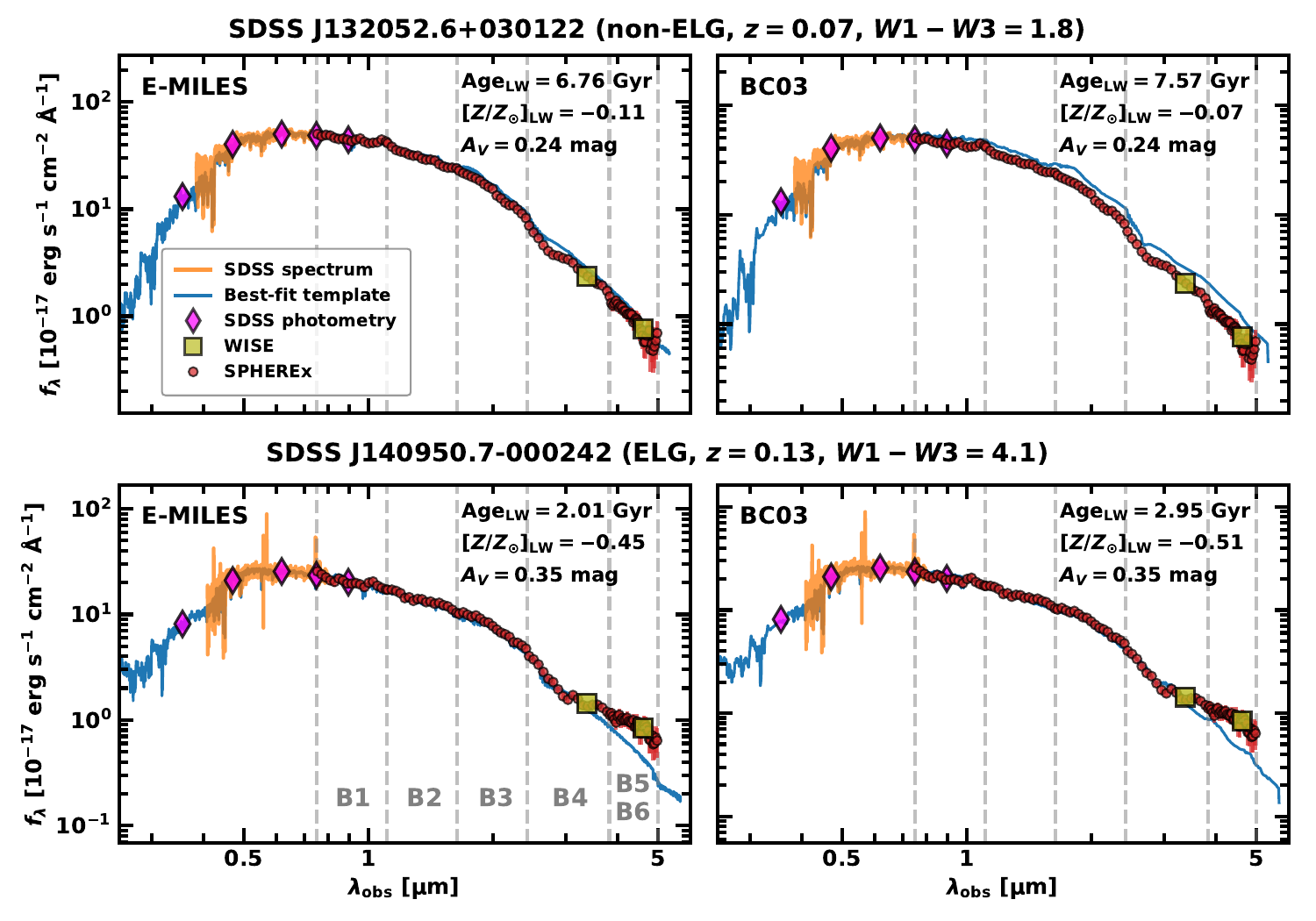}
\caption{
Example SEDs with observational data and modeled SEDs of two representative galaxies: a non-ELG (SDSS J132052.6$+$030122; top two panels) and an ELG (SDSS J140950.7$-$000242; bottom two panels).
Each row displays the best-fit results using two different SPS model sets, E-MILES (left) and BC03 (right).
In each panel, orange and blue solid lines represent the observed SDSS spectra and the best-fit pPXF templates, respectively.
Gray vertical dashed lines denote the wavelength boundaries of SPHEREx bands: $1.11~{\rm \mu m}$ (Bands 1 and 2), $1.64~{\rm \mu m}$ (Bands 2 and 3), $2.42~{\rm \mu m}$ (Bands 3 and 4), and $3.82~{\rm \mu m}$ (Bands 4 and 5/6).
The boundary at $4.42~{\rm \mu m}$ (Bands 5 and 6) is not shown due to narrow wavelength range of these high-resolution bands.
Multiwavelength photometric data are denoted by magenta diamonds (SDSS model magnitudes), yellow squares (WISE all-sky magnitudes), and red circles (SPHEREx QR2 data).
All data are corrected for foreground extinction, with aperture corrections applied to the SDSS spectra.
In these diagrams, the WISE photometric data are multiplied by the scaling factor between SPHEREx and SDSS $z$ band, as done for the SPHEREx data.
The stellar population properties of these galaxies, such as luminosity-weighted age (${\rm Age_{\rm LW}}$), metallicity ($[Z/Z_{\odot}]_{\rm LW}$), and $V$-band extinction ($A_{V}$), are summarized in the upper-right corner of each panel.
\label{fig:sed_full}}
\end{figure*}

\section{Methods}
\label{sec:method}

\subsection{Full-spectrum Fitting Based on SPS Models}
\label{sec:ppxf}

In this study, we adopted four NIR-covering SPS models to investigate the model dependence of NIR SEDs of galaxies: E-MILES \citep{vaz16}, BC03 \citep{bru03}, CB19 \citep{pla19}, and FSPS \citep{con09, con10a}.
The same set of SPS models was examined by \citet{LeeJH25}, where detailed descriptions of each model were provided (see their Section 3).
However, compared to \citet{LeeJH25}, we excluded the SSP templates with stellar ages younger than 70 Myr (${\rm log~Age~(yr) < 7.8}$), as this study focused on integrated galaxy light rather than spatially resolved regions in nearby late-type galaxies.
In addition, in E-MILES, the spectral templates of superyoung stellar populations (${\rm log~Age~(yr) < 7.8}$) do not cover the NIR wavelengths, which can introduce another systematic bias for the comparisons of the modeled NIR SEDs.
Furthermore, including superyoung stellar templates for the other three models made minimal changes to the overall trends.
For these reasons, we decided to restrict the stellar spectral templates to older ages with ${\rm log~Age~(yr) > 7.8}$ in this study.
Except for this restriction, the adopted combinations of age, metallicity, isochrones, and initial mass function \citep{cha03} in the SPS models are identical to those in \citet{LeeJH25}.
The Maraston 2005 model was excluded from our analysis due to its low spectral resolution and weak predictability of SEDs at longer wavelengths than $2.5~{\rm \mu m}$, although it is considered a good comparison model to examine the contribution of TP-AGB stars to NIR SEDs.
We further performed simple tests with different combinations of the initial mass function from \citet{sal55} and the BaSTI isochrones \citep{pie04} modeled in E-MILES, but the general trends of our analysis do not meaningfully change.

To estimate the model-based NIR SEDs, we conducted full-spectrum fitting of the SDSS optical spectra of our sample galaxies.
We used the Penalized Pixel-Fitting method \citep[pPXF;][]{cap23} to derive the stellar population properties (e.g, ages and metallicities) and to find the best-fit SEDs from optical to NIR wavelengths.
We adopted a two-step fitting process, with the first run for determining stellar kinematics and the second for stellar population parameters and best-fit SEDs.
For the pPXF fitting, we shifted the observed spectra to the rest frame using SDSS redshift values and masked emission lines from ionized gas ([\ion{O}{2}]$\lambda\lambda3727,3729$, [\ion{O}{3}]$\lambda\lambda4959,5007$, [\ion{N}{2}]$\lambda\lambda6548,6584$, [\ion{S}{2}]$\lambda\lambda6717,6731$, and Balmer lines) and telluric features ([\ion{O}{1}]$\lambda5577$ and [\ion{O}{1}]$\lambda6300$) to focus on the stellar spectral continuum.
We restricted the fitting wavelength range from $3600~{\rm \AA}$ to $7000~{\rm \AA}$, where the spectra have high signal-to-noise ratios and are less contaminated by sky residuals at redder wavelengths.
The detailed pPXF configurations, including hyperparameters such as \texttt{adeg}, \texttt{mdeg}, \texttt{reddening}, and \texttt{regul}, were the same as those used in \citet{LeeJH25}.

We applied the dust extinction law of \citet{cal00} for all SPS models, which assumes $R_{V}=A_{V}/E(B-V)=4.05$ as a default configuration in pPXF, to minimize potential systematic biases induced from the use of different dust laws.
In addition, we fixed the $A_{V}$ values for all SPS models to those derived from E-MILES.
This choice is reasonable because adopting the same $A_{V}$ across all SPS models can remove the dependence of the SED comparison on differences in extinction magnitude.
Moreover, the $A_{V}$ distributions appear nearly identical among different SPS models when $A_{V}$ is set to a free parameter (see Figure 6 and Table 1 of \citeauthor{LeeJH25} \citeyear{LeeJH25}), indicating that fixing $A_{V}$ may take minimal effects on the spectral fitting results.

Finally, we constructed the best-fit SEDs by combining input model templates with the pPXF-derived weight values.
It is noteworthy that we did not include SPHEREx data in the pPXF fitting procedure.
Through this approach, this study can compare the predicted NIR SEDs purely based on optical spectra with dense NIR observations from SPHEREx.

\subsubsection{Validation of Stellar Kinematics}

Before determining the stellar population parameters and NIR SEDs of the sample galaxies, it is necessary to validate their stellar kinematics---radial velocity ($v_{\rm rad}$) and velocity dispersion ($\sigma_{v}$)---derived from pPXF.
Here we rejected galaxies with large discrepancies in radial velocity ($|\Delta v_{\rm rad}| > 150~{\rm km~s^{-1}}$), unrealistic velocity dispersion values ($\sigma_{v}>400~{\rm km~s^{-1}}$), and exceptionally high chi-square values ($\chi^{2}>100$) from our pPXF results.
These cuts left 4108 galaxies from the initial sample of 4172 galaxies.

{\color{blue} \textbf{Figure \ref{fig:comp_kin}}} compares stellar kinematics derived from this work with the SDSS archival values (\texttt{z} and \texttt{velDisp} from the \texttt{SpecObjAll}).
The median offsets ($\Delta_{\rm med}$) are $8.3~{\rm km~s^{-1}}$ for radial velocity and $-0.6~{\rm km~s^{-1}}$ for velocity dispersion, indicating no significant bias in stellar kinematics measurements in this study.
Furthermore, we did not find any systematic dependence on galaxy type (non-ELGs vs. ELGs).
For radial velocity, the median offsets and scatters are $\Delta_{\rm med}=9.8~{\rm km~s^{-1}}$ and $\sigma_{\rm resid}=12~{\rm km~s^{-1}}$ for non-ELGs, and $\Delta_{\rm med}=2.9~{\rm km~s^{-1}}$ and $\sigma_{\rm resid}=17~{\rm km~s^{-1}}$ for ELGs.
In the comparison of velocity dispersion, the two galaxy populations show a subtle difference in their median offsets, with $\Delta_{\rm med}=-5.9~{\rm km~s^{-1}}$ for ELGs ($\langle\sigma_{v}\rangle\sim90~{\rm km~s^{-1}}$) and $\Delta_{\rm med}=1.2~{\rm km~s^{-1}}$ for non-ELGs ($\langle\sigma_{v}\rangle\sim150~{\rm km~s^{-1}}$), with weak positive trends notable in the right panel of {\color{blue} \textbf{Figure \ref{fig:comp_kin}}} (see also Figure 1 in \citeauthor{tho13} \citeyear{tho13}).
In addition, ELGs exhibit a larger scatter in velocity dispersion ($\sigma_{\rm resid}=41~{\rm km~s^{-1}}$) compared to non-ELGs ($\sigma_{\rm resid}=12~{\rm km~s^{-1}}$), suggestive of possible uncertainties in their measurements.
These uncertainties may be related to the systematic differences in the detailed methods between the SDSS pipeline and pPXF, such as spectral resolution of templates, fitting wavelength range, or optimizing algorithms.
Overall, the comparison of stellar kinematics demonstrates good agreement between our measurements and SDSS values, confirming the reliability of the stellar kinematics and enabling robust determination of stellar population properties and NIR SEDs.

\subsection{Calculating the Weighted Offset}
\label{sec:offset}

In this section, we explain how we quantified the relative differences between the modeled SEDs and SPHEREx observations.
{\color{blue} \textbf{Figure \ref{fig:sed_full}}} displays the examples of observed and modeled SEDs for a non-ELG (upper panels) and an ELG (lower panels), derived using the two different SPS models of E-MILES (left panels) and BC03 (right panels).
Although our sample galaxies are selected to be compact, their total light is not fully captured within the $3''$-diameter SDSS spectroscopic fibers.
To account for this aperture effect, we computed the ratio of the SDSS model flux to the fiber flux in the $r$ band and multiplied both the SDSS spectra (orange lines) and the best-fit model SEDs (blue lines) by this factor.

To account for the photometric discrepancy between SDSS and SPHEREx arising from differing PSFs and photometric methods, we applied an additional scaling factor to the observed SPHEREx data. This factor was calculated from the ratio of the SDSS $z$-band magnitude to the synthetic $z$-band magnitude derived from the SPHEREx data using \texttt{pyphot}.
Here, we used the $z$ band for this calibration because its wavelength range is fully covered by the SPHEREx channels.

We found that the original SPHEREx fluxes were only marginally fainter than the SDSS $z$-band fluxes by a median offset of $\sim 4\%$.
This is likely because the SDSS model light profiles of sample galaxies still tend to be slightly extended compared to the SPHEREx PSF profile.
Indeed, these flux offsets show a weak correlation with apparent sizes of galaxies, with a Spearman coefficient of $r_{S}=0.31~(p<0.05)$ with respect to the Petrosian radius at SDSS $r$ band.
However, the range of galaxy sizes is relatively narrow ($\sim1.5-3~{\rm arcsec}$ for Petrosian radius) due to our initial selection for compact sizes ($\texttt{petroRad}<3.1~{\rm arcsec}$), leading to $\sim95\%$ of the sample galaxies having lower SPHEREx-to-SDSS offsets than $20\%$.
Applying the SPHEREx-to-SDSS flux scaling minimizes the systematic offsets arising from instrumental differences between SDSS and SPHEREx.
It is important to note that this correction does not affect the overall shape of observed SPHEREx SEDs, implying that the wavelength-dependent trends of the SED differences do not change.
For subsequent analysis, we excluded galaxies with SPHEREx-to-SDSS offsets greater than $20\%$, to account for possible inaccuracies in SPHEREx or SDSS model photometry.
This cross-calibration resulted in our final analysis sample of 3,889 galaxies among the 4,066 galaxies with valid SPHEREx data points.
Using these procedures, both the observed and modeled SEDs of the sample galaxies were consistently constructed across the optical to NIR wavelength range.

Meanwhile, no additional flux scaling was required for the WISE photometry because we only used the WISE ${\rm W1-W3}$ color in our analysis.
The WISE W1 photometry, where the wavelength range is fully overlapped with SPHEREx, shows median fluxes only $\sim3.5\%$ lower than the synthetic W1 fluxes derived from the original SPHEREx photometry.
This small difference indicates that the SPHEREx and WISE photometry are already in good agreement for our compact galaxy sample.

In {\color{blue} \textbf{Figure \ref{fig:sed_full}}}, the best-fit templates from E-MILES and BC03 exhibit subtle differences, particularly in SPHEREx Bands 4 to 6 (in the observer frame for this figure).
To statistically compute these systematic differences between the modeled and observed SEDs, we first calculated the relative difference ($\Delta f$) between the model-expected SEDs from pPXF ($f_{\rm model}$) and the observed SPHEREx data ($f_{\rm SPHEREx}$):
\begin{equation}
    \Delta f=\frac{f_{\rm model}-f_{\rm SPHEREx}}{f_{\rm SPHEREx}}
\end{equation}
Then, we derived the weighted offset parameter ($\mu_{\Delta}$) over the NIR wavelengths defined as follows:
\begin{equation}
    \mu_{\Delta}=\frac{\sum_{i} w_i \Delta f_i}{\sum_{i} w_i}
\end{equation}
where $i$ denotes the $i$-th wavelength bin for SPHEREx.
In this calculation, the wavelengths and fluxes are converted into the rest frame using the redshift information of each galaxy sample.
The weights ($w_i$) are calculated as the inverse square of the uncertainties in $\Delta f_i$, propagated with the measured flux uncertainties for the SPHEREx fluxes and an uncertainty floor of $10\%$ for the modeled fluxes.
Here, we adopted the magnitude unit for the weighted offset parameter.
We set positive values ($\mu_{\Delta}>0$) to indicate that the models predict brighter fluxes than SPHEREx observations and negative values ($\mu_{\Delta}<0$) for the opposite.
Using this metric, we evaluate systematic differences in NIR SEDs across SPS model selections.

\begin{figure*}
\centering
\includegraphics[width=\textwidth]{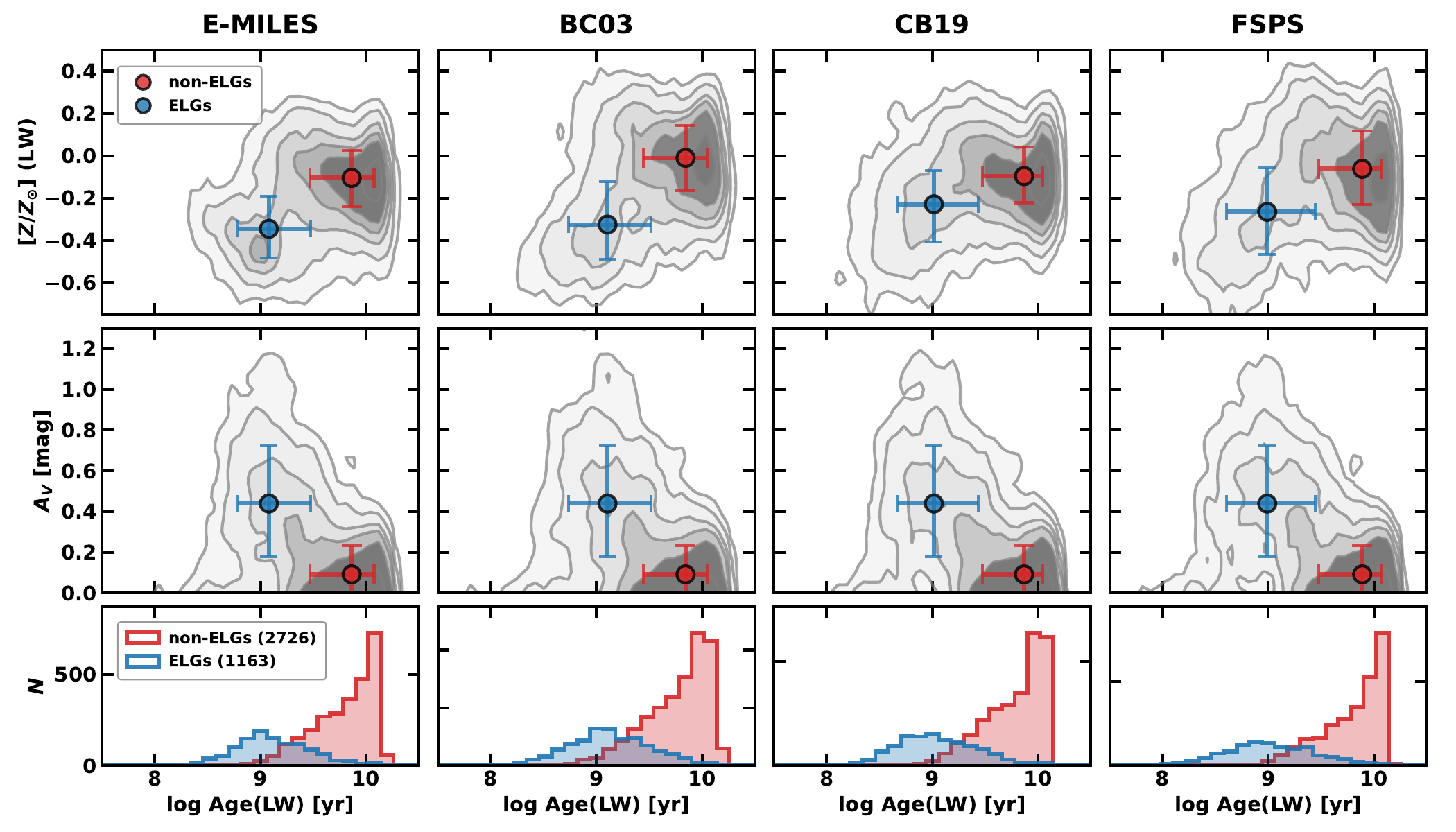}
\caption{
Distributions of stellar population properties derived from pPXF for different SPS models using SDSS spectra.
Each column represents the results from the four different SPS models: E-MILES, BC03, CB19, and FSPS (left to right).
In the top and middle rows, the luminosity-weighted metallicity and the $V$-band extinction magnitude of the sample galaxies are displayed as a function of their luminosity-weighted stellar age.
Gray contours indicate the 50th, 65th, 80th, 90th, 95th, and 99th percentiles of the smoothed two-dimensional number density of galaxies.
The median values of the properties are denoted by red (non-ELGs) and blue (ELGs) circles, with error bars corresponding to the 16th to 84th percentile ranges.
The bottom row presents the age distributions of non-ELGs (red) and ELGs (blue) as histograms for each SPS model.
\label{fig:dist2}}
\end{figure*}

\section{Results}
\label{sec:result}

\subsection{Distributions of Galaxy Properties}
\label{sec:res1}

In this section, we present the stellar population properties (i.e., stellar age, metallicity, and $A_{V}$) of the final sample of 3,889 compact galaxies.
We also examine the distribution of observed WISE colors for non-ELGs and ELGs, which serve as an effective indicator of non-stellar contributions in the NIR.

\subsubsection{Stellar Population Properties}

{\color{blue} \textbf{Figure \ref{fig:dist2}}} shows the distributions of (luminosity-weighted) stellar ages, metallicities, and $A_{V}$ for different SPS models.
As expected, most of the sample galaxies show old stellar ages (${\rm log~Age~(yr) > 9.5}$), which are consistent with $\sim70\%$ of the sample being non-ELGs with evolved stellar populations.
In contrast, ELGs show much younger stellar ages, with nearly symmetric distributions centered around $\sim 1~{\rm Gyr}$.
These two populations also exhibit clear differences in stellar metallicity and $A_{V}$, with non-ELGs showing notably higher metallicities and lower $A_{V}$ compared to ELGs.
This trend is a natural consequence: non-ELGs predominantly consist of evolved stars with quenched star formation, which is accompanied by enriched chemical elements and deficient interstellar medium (ISM), whereas ELGs show ongoing star formation activity and more abundant ISM ingredients.
These full-spectrum fitting results demonstrate that our compact galaxy sample spans a wide range of stellar population properties, with no significant selection bias for the NIR SED comparison performed in this study.

The overall distributions of these stellar population parameters seem to be broadly consistent across the SPS models.
The age distributions are nearly identical among the models for both non-ELGs and ELGs, mitigating the impact of age-dependent effects on the SED comparison analysis.
There are subtle differences in the metallicity distributions for different SPS models.
For ELGs, the E-MILES and BC03 models predict median metallicities lower by $\sim0.1~{\rm dex}$ compared to CB19 and FSPS, whereas this offset is absent among non-ELGs.
A similar discrepancy was reported in \citet{LeeJH25} for spatially resolved properties of late-type galaxies, which have analogous properties with ELGs in this study.
In particular, the BC03 model predicts the widest spread of metallicities between non-ELGs and ELGs ($\sim0.33~{\rm dex}$ in median), while other models yield $\sim0.15-0.2~{\rm dex}$.
This behavior is also consistent with \citet{LeeJH25}, presenting large metallicity separations in BC03.
Overall, these results suggest that the composite stellar population properties are relatively insensitive to the choice of SPS models and are not likely to introduce significant degeneracies in the SED comparison.

\begin{figure}
\centering
\includegraphics[width=0.47\textwidth]{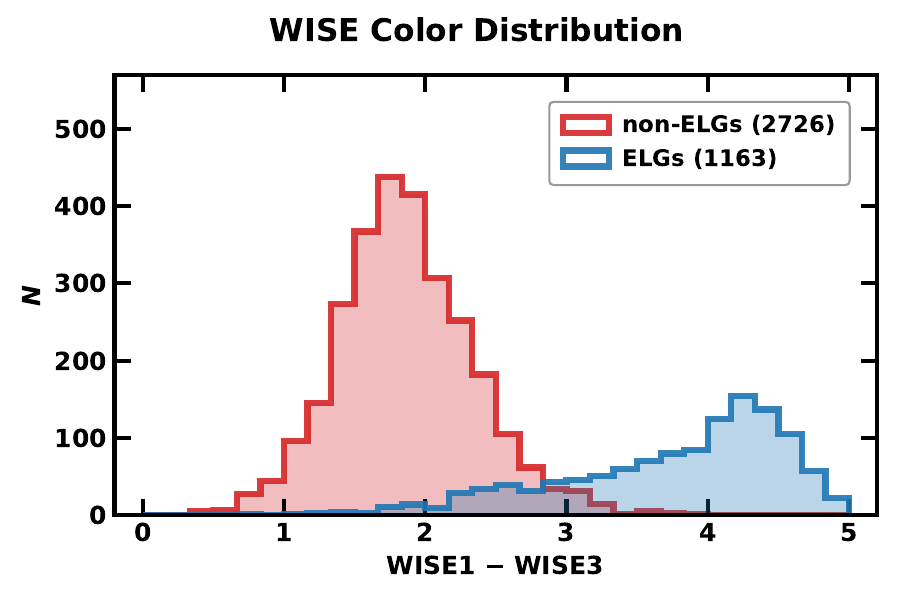}
\caption{
Histograms representing the WISE color (${\rm W1-W3}$) distributions of non-ELGs (red) and ELGs (blue).
\label{fig:wise}}
\end{figure}

\subsubsection{WISE $W1-W3$ Colors}

In {\color{blue} \textbf{Figure \ref{fig:wise}}}, we present the WISE ${\rm W1-W3}$ color distributions for non-ELGs and ELGs.
The two populations exhibit a clear bimodal distribution, with median values of ${\rm W1-W3}=1.8$ for non-ELGs and ${\rm W1-W3}=4.0$ for ELGs.
This indicates that ELGs exhibit a higher contribution from dust emission in the MIR, leading to significantly brighter W3 band luminosities than non-ELGs, which are expected to have less ISM dust.
This clear separation in WISE color demonstrates that optical emission-line properties are closely linked to infrared photometric properties, and that the ${\rm W1-W3}$ color provides a useful observational proxy for assessing star formation activity of galaxies and non-stellar contributions in their NIR SEDs.
Using this WISE color, we further investigate its relation to the weighted offset of NIR SEDs in {\color{blue} \textbf{Section \ref{sec:res2_2}}}.

\subsection{Offsets in NIR SEDs from the SPS Models}
\label{sec:res2}

\begin{figure}
\centering
\includegraphics[width=0.5\textwidth]{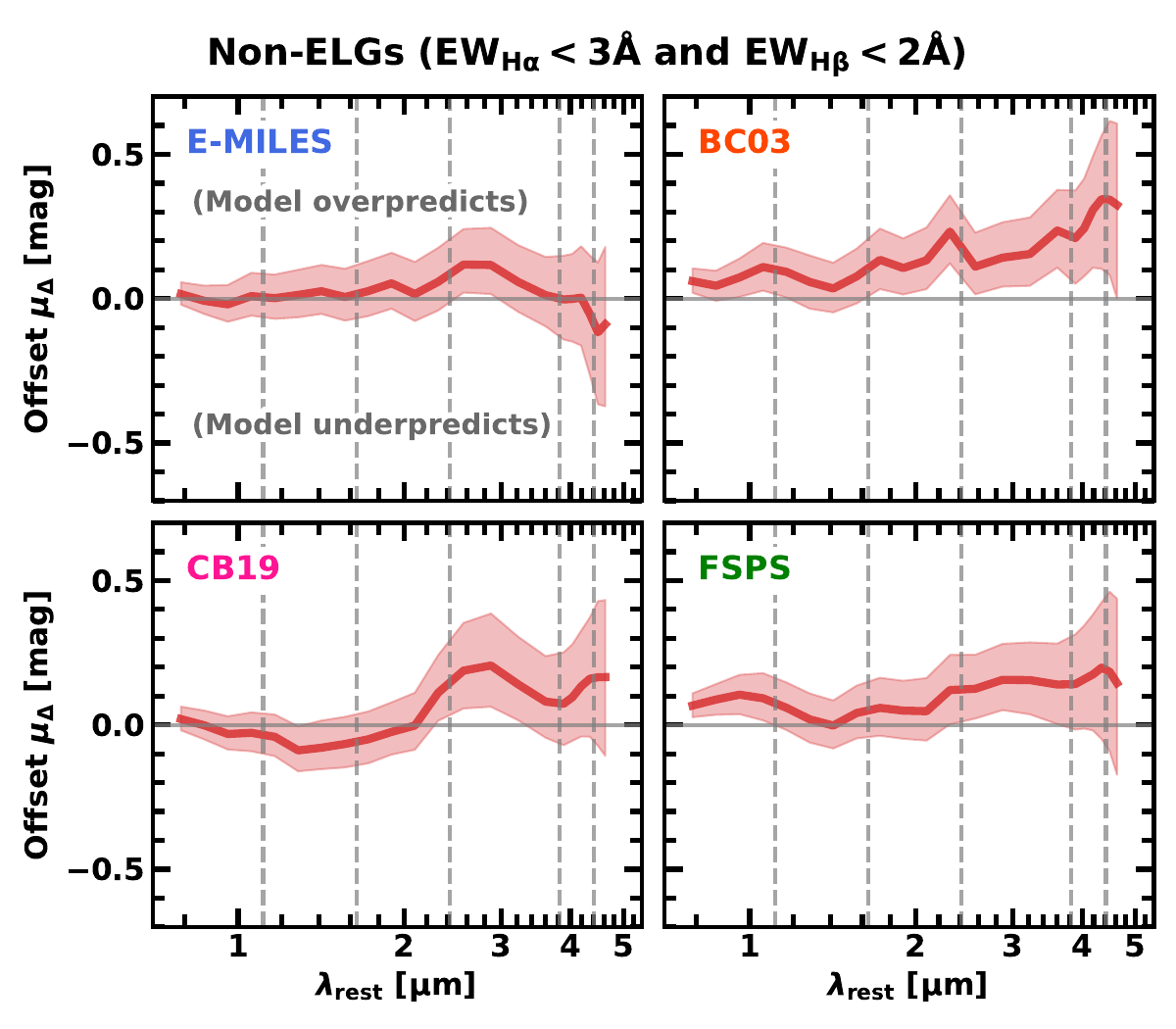}
\caption{
Distributions of the weighted offsets ($\mu_{\Delta}$) between the modeled SEDs and SPHEREx observations for non-ELGs, shown as a function of rest-frame NIR wavelength.
Each panel represents the results for a different SPS model.
Here $\mu_{\Delta}>0$ indicates that the modeled NIR fluxes are brighter than those observed by SPHEREx, while $\mu_{\Delta}<0$ indicates that the modeled fluxes are fainter.
In each panel, the median $\mu_{\Delta}$ values are represented by red thick curves, and the 16th to 84th percentiles are represented by red shaded regions.
The $\mu_{\Delta}=0$ level is marked with gray horizontal lines for reference.
Gray vertical lines mark the wavelength boundaries between SPHEREx bands, as a guide to the rest-frame wavelength intervals.
\label{fig:offset_nE}}
\end{figure}

We investigate the weighted offsets ($\mu_{\Delta}$) calculated in {\color{blue} \textbf{Section \ref{sec:offset}}} depending on SPS model, SPHEREx band, and galaxy type.
We present the distributions of $\mu_{\Delta}$ separately for non-ELGs ({\color{blue} \textbf{Figure \ref{fig:offset_nE}}}) and ELGs ({\color{blue} \textbf{Figure \ref{fig:offset_yE}}}), as their SED offsets are expected to diverge significantly at longer wavelengths ($\lambda \gtrsim 2.4~{\rm \mu m}$) due to different non-stellar contributions.
Among the two galaxy populations, non-ELGs provide a more suitable sample for investigating the dependence of stellar NIR SEDs on the adopted SPS model because of their weaker contributions from non-stellar emission.

\subsubsection{Overall Trends of the SED Offsets}
\label{sec:res2_1}

In {\color{blue} \textbf{Figure \ref{fig:offset_nE}}}, the E-MILES model yields the closest agreement with the SPHEREx observations across all bands.
The median offsets are minimal across the NIR wavelengths, but a slight offset of $\sim0.1~{\rm mag}$ is noted at $\lambda_{\rm rest} = 2.4-3.2~{\rm \mu m}$, corresponding to $\sim10\%$ brighter SED predictions.
For the other SPS models, similar but stronger trends of overprediction appear at $\lambda_{\rm rest} = 2.4-3.2~{\rm \mu m}$, with median $\mu_{\Delta}\sim0.2~{\rm mag}$ (i.e., $\sim20\%$ brighter SEDs).
At the SPHEREx high-resolution bands ($\lambda_{\rm rest} = 3.8-5.0~{\rm \mu m}$), despite large variances in $\mu_{\Delta}$ due to the low number of rest-frame data points and large photometric uncertainties, the general trends of median $\mu_{\Delta}$ are still model-dependent.
Therefore, the SPHEREx uncertainties cannot solely explain the systematic difference in NIR SEDs at these long wavelengths.

Regarding the detailed behavior of $\mu_{\Delta}$, there are several different trends among the SPS models.
First, while the E-MILES model yields reduced offsets at $\lambda_{\rm rest} = 3.8-5.0~{\rm \mu m}$, the other models continue to overpredict the SEDs at these wavelengths by $0.15-0.30~{\rm mag}$.
This indicates that these models tend to predict $>15\%$ brighter SEDs than E-MILES.
In particular, the BC03 model shows the largest discrepancy of $\sim0.3~{\rm mag}$ relative to E-MILES.
Second, at shorter wavelengths ($\lambda_{\rm rest}<2.4~{\rm \mu m}$), the BC03 and FSPS models predict slightly brighter SEDs by $\sim0.1~{\rm mag}$ compared to E-MILES and CB19.
This difference is already apparent at $\lambda_{\rm rest}<1.1~{\rm \mu m}$, where E-MILES and CB19 have negligible offsets but BC03 and FSPS show median $\mu_{\Delta}$ of $0.07$ to $0.09~{\rm mag}$.
Such small discrepancies may partially arise from uncertainties in the SPHEREx flux calibration process done in {\color{blue} \textbf{Section \ref{sec:offset}}}.
Notably, the median $\mu_{\Delta}$ values from BC03 steadily increase from $\lambda_{\rm rest}=0.8~{\rm \mu m}$ to $2.4~{\rm \mu m}$, while those from the other models remain nearly constant.
This suggests a systematic discrepancy of BC03 relative to the other models.
Based on these results of non-ELGs, we discuss the implications of stellar emissions from SPHEREx and SPS models in {\color{blue} \textbf{Section \ref{sec:discuss}}}.

\begin{figure}
\centering
\includegraphics[width=0.5\textwidth]{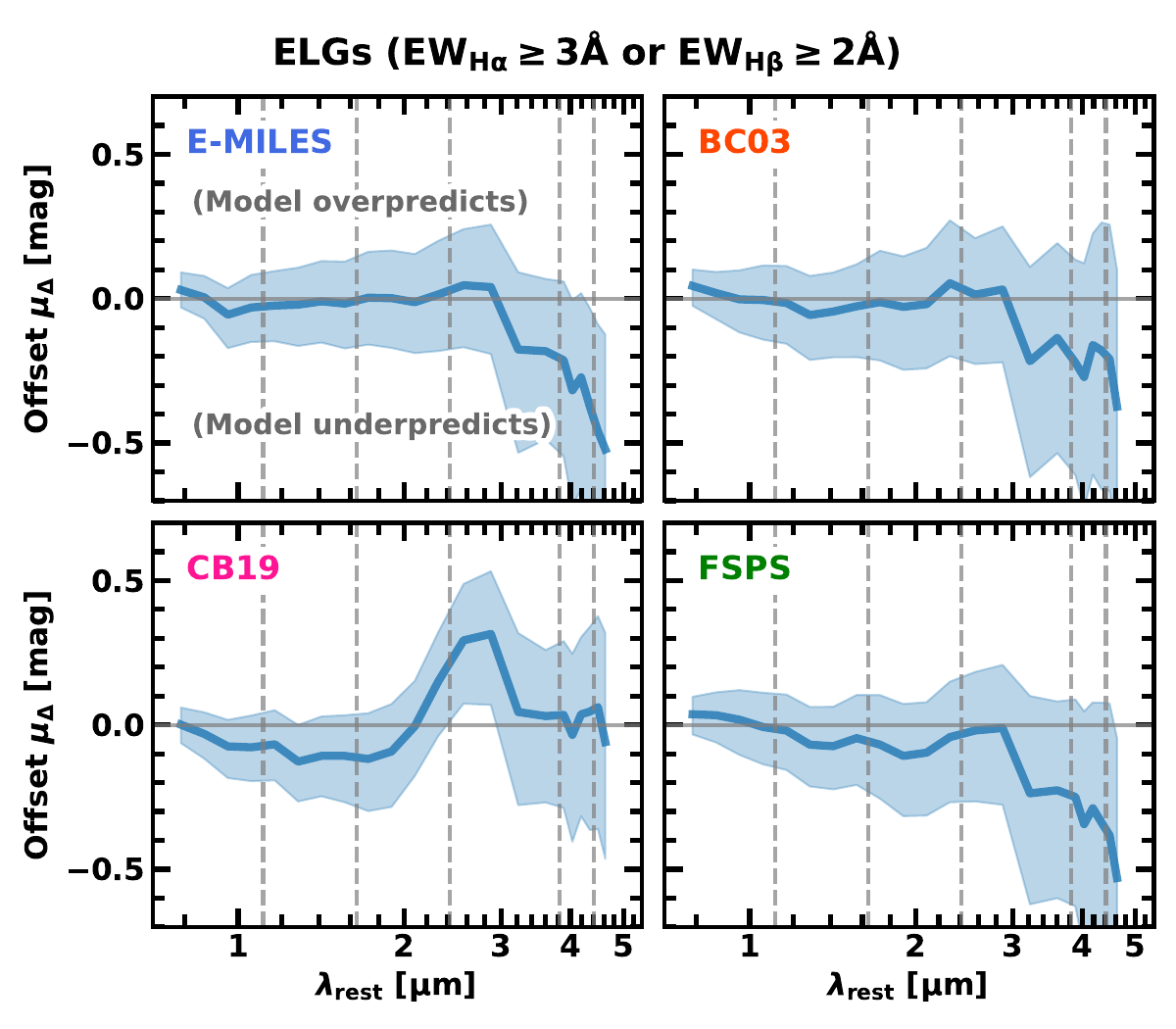}
\caption{
Same as {\color{blue} \textbf{Figure \ref{fig:offset_nE}}}, but for ELGs.
\label{fig:offset_yE}}
\end{figure}

For ELGs, as a control sample including non-stellar emissions, {\color{blue} \textbf{Figure \ref{fig:offset_yE}}} represents their offset distributions.
Since ELGs contain substantial contributions of non-stellar emission, direct comparisons between stellar model SEDs and SPHEREx observations cannot be solely interpreted as model dependence at $\lambda_{\rm rest} = 3.8-5.0~{\rm \mu m}$.
Across all wavelengths, ELGs exhibit larger deviation in $\mu_{\Delta}$ than non-ELGs.
At $\lambda_{\rm rest} < 2.4~{\rm \mu m}$, all SPS models show median offsets smaller than $0.1~{\rm mag}$, indicative of broad agreement with the observations.

At longer wavelengths ($\lambda_{\rm rest} > 2.4~{\rm \mu m}$), the modeled SEDs tend to be increasingly underpredicted relative to the observations, due to the large contribution of dust emission detected by SPHEREx.
Although these offsets do not primarily arise from the stellar SED prediction from SPS models, the amplitudes of $\mu_{\Delta}$ vary with the models.
Surprisingly, the CB19 model rather yields $\mu_{\Delta}>0$, indicating a relative overprediction of the NIR SEDs even in these dust-dominated wavelength ranges.

To summarize, for non-ELGs, all SPS models show a clear overprediction ($\mu_\Delta>0$) of $\sim0.1-0.2~{\rm mag}$ at $\lambda_{\rm rest}=2.4-3.2~{\rm \mu m}$.
This SED overprediction persists at $\lambda_{\rm rest}=3.8-5.0~{\rm \mu m}$ for all models except E-MILES.
For ELGs, this trend is reversed to underprediction ($\mu_\Delta<0$) at $\lambda_{\rm rest}=3.8-5.0~{\rm \mu m}$ due to significant dust contributions, although CB19 still yields positive median values of $\mu_\Delta$.

\begin{figure*}
\centering
\includegraphics[width=\textwidth]{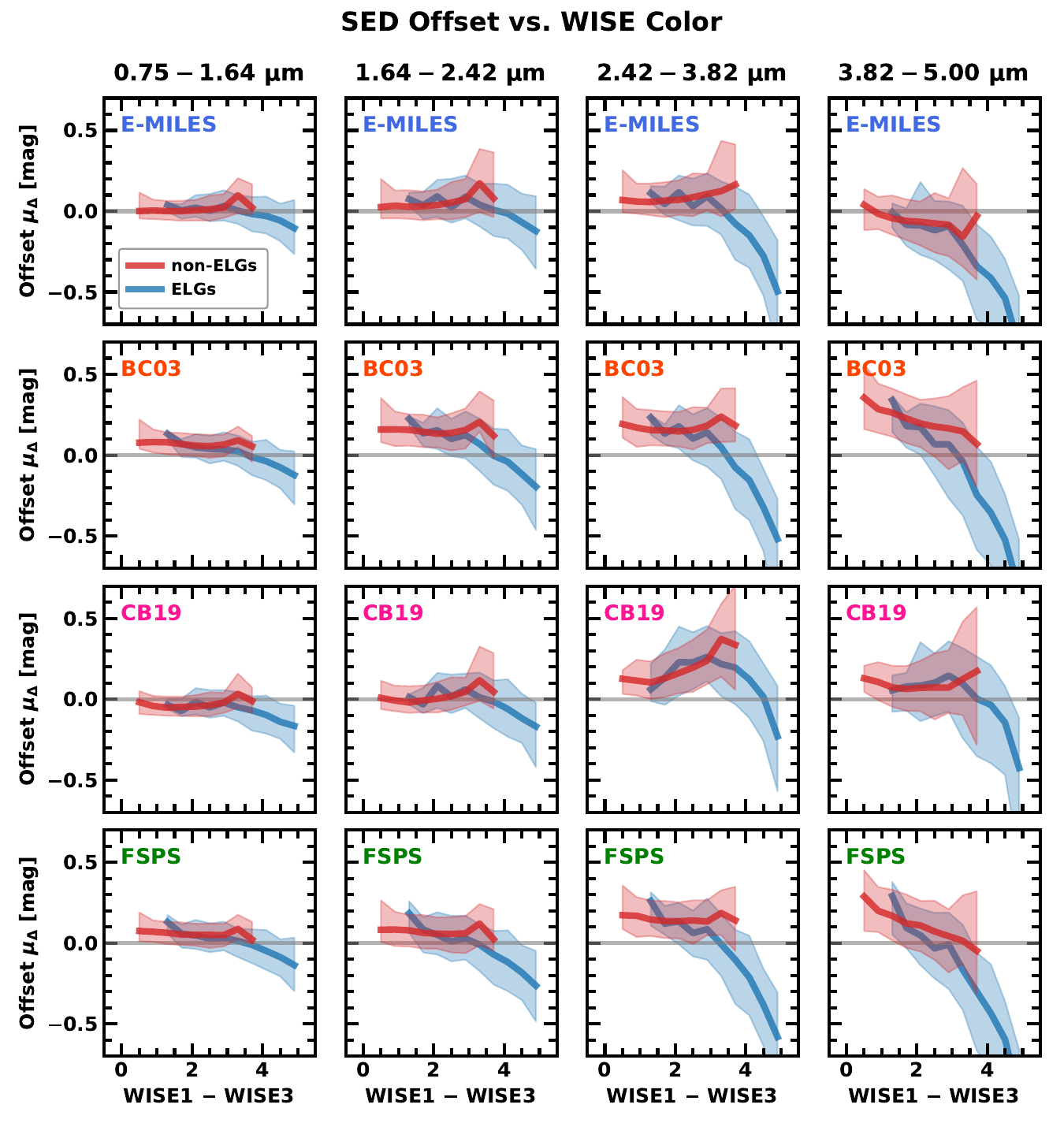}
\caption{
Offset distributions for non-ELGs and ELGs as a function of WISE ${\rm W1-W3}$ color.
From top to bottom, each row represents the results based on E-MILES, BC03, CB19, and FSPS, respectively.
For each model, the columns display the relations for rest-frame wavelength bins: $\lambda_{\rm rest}=0.75-1.64~{\rm \mu m}$ (first column), $\lambda_{\rm rest}=1.64-2.42~{\rm \mu m}$ (second), $\lambda_{\rm rest}=2.42-3.82~{\rm \mu m}$ (third), and $\lambda_{\rm rest}=3.82-5.00~{\rm \mu m}$ (fourth).
In each panel, we plot the median values and the 16th--84th percentile ranges of non-ELGs (red curves and shaded regions) and ELGs (blue curves and shaded regions) for the color bins.
\label{fig:offset_w13}}
\end{figure*}

\subsubsection{Trends of the SED Offsets with WISE Color}
\label{sec:res2_2}

We further investigate the relationship between the SED offsets and WISE color to disentangle the non-stellar contribution from the overall trends shown in {\color{blue} \textbf{Section \ref{sec:res2_1}}}.
{\color{blue} \textbf{Figure \ref{fig:offset_w13}}} depicts the relationship between $\mu_{\Delta}$ and WISE ${\rm W1-W3}$ color in four rest-frame wavelength bins.

As expected, all SPS models consistently exhibit the SED underprediction across all wavelength bins in the red-color regime (${\rm W1-W3} \gtrsim 3$), where most ELGs are located.
In particular, strong negative correlations appear at $\lambda_{\rm rest}>2.42~{\rm \mu m}$, which are predominantly affected by dust emission (e.g., the PAH feature at $3.3~{\rm \mu m}$).
Overall, the trends with ${\rm W1-W3}$ colors are in agreement between non-ELGs and ELGs, suggesting that the WISE color provides a more direct indicator for dust contribution rather than galaxy type classified by optical emission-line features.
Therefore, the WISE color can provide a useful observational proxy for revealing the increasing contribution of non-stellar emission to the SED offsets at $\lambda_{\rm rest}>2.42~{\rm \mu m}$, particularly for ELGs.

Meanwhile, non-ELGs show no significant correlation between the SED offsets and ${\rm W1-W3}$ color across all wavelengths.
This indicates that the SED offsets of non-ELGs observed at $\lambda_{\rm rest}>2.42~{\rm \mu m}$ are mainly driven by differences in the stellar emission predicted from SPS models rather than by variations in dust emission.
From the age--$\mu_{\Delta}$ relation presented in {\color{blue} \textbf{Section \ref{sec:dis1}}}, we can determine the underlying stellar populations responsible for the SED offsets in non-ELGs.

Conclusively, ELGs with young stellar populations (${\rm log~Age~(yr) \lesssim 9.0}$) or red WISE colors (${\rm W1-W3} \gtrsim 3$) exhibit substantial dust emission at $\lambda \gtrsim 2.4~{\rm \mu m}$, making them less suitable for distinguishing model-dependent variations in the stellar NIR SEDs predicted by SPS models.
Nevertheless, presenting the SED offsets of ELGs in this study remains valuable, as the dense spectra coverage of SPHEREx enables us to effectively quantify the contribution of dust emission to their NIR SEDs.
In the following discussion of SPS model dependence, we primarily focus on non-ELG populations to assess the performance of SPS models in predicting stellar NIR SEDs.

\section{Discussion}
\label{sec:discuss}

\subsection{Model-dependent Relations between SED Offset and Stellar Population}
\label{sec:dis1}

In this section, we examine how the SED offsets between SPS models and SPHEREx observations correlate with galaxy properties (i.e., stellar age and metallicity) and the choice of SPS models.
As the overall distributions of age and metallicity appear similar among the SPS models, we set the parameters from E-MILES as the reference for dividing subsamples in the following discussion.
We regard non-ELGs as the main analysis sample throughout the discussion to purely compare stellar emissions between models and observations.
The $V$-band extinction ($A_{V}$) could affect the SED offsets in principle, but it does not contribute to the model-dependent trends of SED offsets because $A_{V}$ was fixed to the values derived from E-MILES in this study.
As a supplementary test, we also examined potential correlations with galaxy redshift and found no significant relationships with the NIR SED offsets of non-ELGs.
Through this analysis, we aim to specify the detailed stellar population properties and the model ingredients responsible for the model-dependent trends with $\mu_{\Delta}$ observed in {\color{blue} \textbf{Section \ref{sec:res2}}}.

\begin{figure*}
\centering
\includegraphics[width=\textwidth]{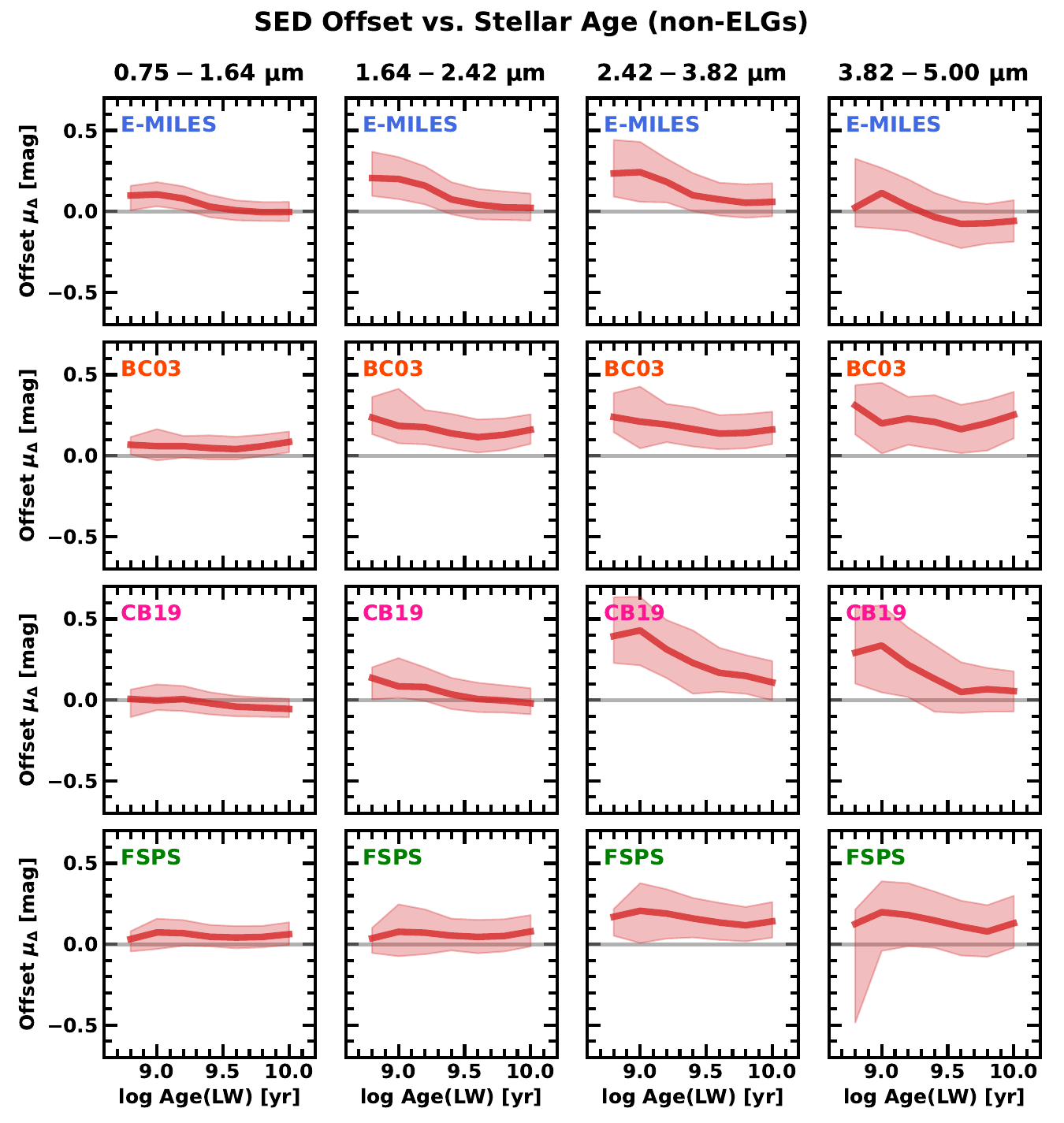}
\caption{
Offset distributions for non-ELGs as a function of stellar age.
From top to bottom, each row represents the results based on E-MILES, BC03, CB19, and FSPS, respectively.
For each model, the columns display the relations for SPHEREx bands in the rest frame: $\lambda_{\rm rest}=0.75-1.64~{\rm \mu m}$ (first column), $\lambda_{\rm rest}=1.62-2.42~{\rm \mu m}$ (second), $\lambda_{\rm rest}=2.42-3.82~{\rm \mu m}$ (third), and $\lambda_{\rm rest}=3.82-5.00~{\rm \mu m}$ (fourth).
In each panel, we plot the median values and the 16th--84th percentile ranges (red curves and shaded regions) for the age bins.
\label{fig:offset_age}}
\end{figure*}

{\color{blue} \textbf{Figure \ref{fig:offset_age}}} illustrates the distribution of $\mu_{\Delta}$ as a function of stellar age.
In this figure, we plot the columns organized by rest-frame wavelength.
At $\lambda_{\rm rest}=0.75-1.64~{\rm \mu m}$, all SPS models produce SEDs broadly consistent with SPHEREx across all age ranges.
The BC03 and FSPS models yield slight overpredictions of $\sim0.1~{\rm mag}$ for old stellar populations (${\rm log~Age~(yr)} > 9.5$), which may explain the small offsets noted in {\color{blue} \textbf{Figure \ref{fig:offset_nE}}}.
In the intermediate-age regime (${\rm log~Age~(yr) \sim 9.0-9.5}$), the offsets from E-MILES show a weak negative correlation with age, leading to overpredictions by $\sim0.1~{\rm mag}$ for galaxies with ${\rm log~Age~(yr) \sim 9}$ at these wavelengths.

The model dependence starts to appear from $\lambda_{\rm rest}=1.64-2.42~{\rm \mu m}$ and becomes more evident at $\lambda_{\rm rest}=2.42-5.00~{\rm \mu m}$.
The E-MILES model tends to show clear overpredictions at $\lambda_{\rm rest}=1.64-3.82~{\rm \mu m}$ for intermediate-age populations, but the offsets are alleviated towards older ages.
At longer wavelengths ($\lambda_{\rm rest}=3.82-5.00~{\rm \mu m}$), the modeled SEDs from E-MILES show better agreement with SPHEREx than those from other models.
The BC03 and FSPS models similarly show overpredictions at the wavelength bins with $\lambda_{\rm rest}>2.42~{\rm \mu m}$, but their offsets have no correlations with stellar age.
On the other hand, the CB19 model shows minimal offset at $\lambda_{\rm rest}<2.42~{\rm \mu m}$ but yields the largest overpredictions at $\lambda_{\rm rest}>2.42~{\rm \mu m}$.
This CB19 trend is likely associated with its steep age--$M/L$ relation in the NIR, as shown in Figure 10 of \citet{LeeJH25}, resulting in lower mass-to-light ratios and consequently brighter luminosities at given stellar mass for younger stellar populations.
These age-dependent effects collectively explain the offset trends of non-ELGs shown in {\color{blue} \textbf{Figure \ref{fig:offset_nE}}}, suggesting that intermediate stellar populations (${\rm log~Age~(yr) < 9.5}$) mainly contribute to larger SED offsets at $\lambda_{\rm rest}>2.42~{\rm \mu m}$ compared to older stellar populations (${\rm log~Age~(yr) \geq 9.5}$) (see {\color{blue} \textbf{Appendix A}}).

\begin{figure*}
\centering
\includegraphics[width=\textwidth]{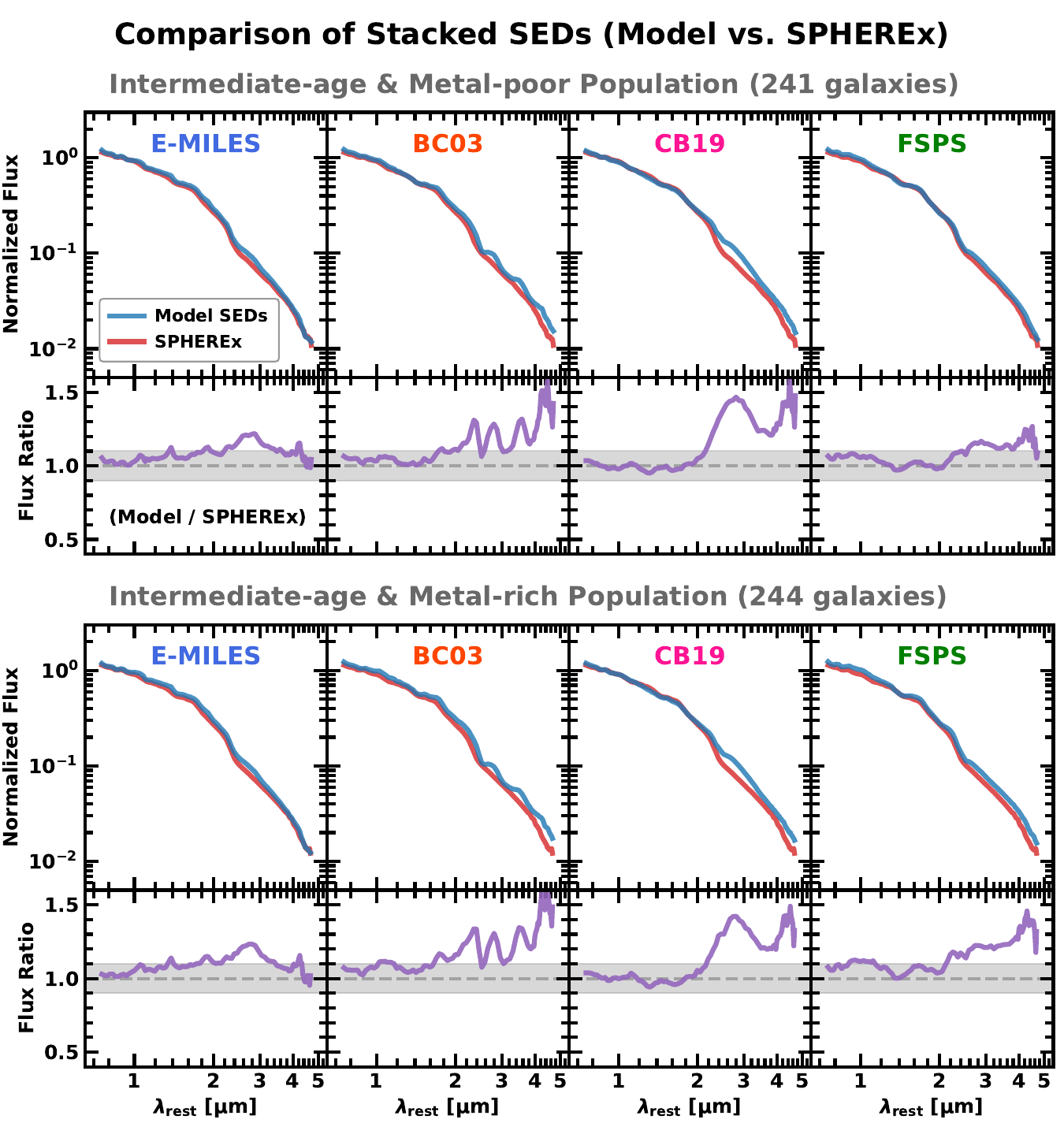}
\caption{
Comparisons of stacked NIR SEDs from SPS models and SPHEREx for intermediate-age stellar populations (${\rm log~Age~(yr)}<9.5$).
The figure is divided into two main panels: the top panel for metal-poor galaxies and the bottom for metal-rich galaxies.
The metallicity bins are defined using the median metallicity of the intermediate-age non-ELGs derived from E-MILES ($[Z/Z_{\odot}]=-0.027$).
Within each main panel, the four columns present the stacked SEDs from E-MILES, BC03, CB19, and FSPS.
In each column, the upper subpanel compares the stacked SEDs from the corresponding SPS model (blue solid line) with those from SPHEREx (red solid line), while the lower subpanel shows the ratio of the model flux to the SPHEREx flux.
The gray-shaded region in the lower subpanels indicates the $10\%$ uncertainty range for reference.
\label{fig:stack}}
\end{figure*}

Focusing on intermediate-age stellar populations, a direct comparison of stacked NIR SEDs between SPS models and SPHEREx observations provides important insights into the specific wavelength ranges responsible for the observed SED offsets.
{\color{blue} \textbf{Figure \ref{fig:stack}}} presents the median stacked NIR SEDs of 485 non-ELGs with ${\rm log~Age~(yr)<9.5}$ derived from the SPS models and SPHEREx data, using the same rest-frame wavelength bins as the 102 SPHEREx spectral channels.
We divide these galaxies into metal-poor (241 galaxies) and metal-rich (244 galaxies) subsamples based on the median metallicity derived from E-MILES ($[Z/Z_{\odot}]=-0.027$) to investigate whether metallicity is an important driver of the SED offsets in intermediate-age non-ELGs.

All SPS models exhibit SED offsets higher than $\sim10\%$ at $\lambda_{\rm rest} \gtrsim 2~{\rm \mu m}$, although the detailed spectral features contributing to these offsets vary with the models.
For E-MILES, the overprediction peaks at $\lambda_{\rm rest}=2.6-3~{\rm \mu m}$, suggesting that the broad absorption features observed by SPHEREx may be underestimated by $\sim 20\%$ at these wavelengths.
The BC03 and FSPS models generally show similar overprediction patterns, but BC03 shows several broad absorption features at $\lambda_{\rm rest} \sim 2.4~{\rm \mu m}$, $2.8~{\rm \mu m}$, and $3.5~{\rm \mu m}$, which produce local minima in the SED offsets, whereas these features are nearly invisible in FSPS.
The BC03 SSP templates exhibit these absorption features primarily at intermediate ages (${\rm log~Age~(yr) \sim 8.5-9.5}$), suggesting that they may be associated with molecular absorption, such as CO or ${\rm H_{2}O}$, in the atmospheres of TP-AGB stars or other cool giants.
CB19 also overpredicts the SED over similar wavelength range ($\lambda_{\rm rest}=2.6-3~{\rm \mu m}$) to E-MILES, but the amplitude is significantly larger, reaching up to $\sim40\%$.
These differences among the NIR SED predictions likely arise from the combined effects of model ingredients including the adopted stellar spectral libraries, isochrones, and treatments of the TP-AGB phase. Specifically, E-MILES in this study uses the empirical IRTF stellar library with Padova isochrones, while the other models are mainly based on the theoretical BaSeL library combined with different stellar evolutionary tracks and TP-AGB models. BC03 adopts the Padova isochrone supplemented with the TP-AGB model of \citet{vas93}, CB19 adopts the PARSEC stellar evolutionary tracks with updated COLIBRI code for TP-AGB phase \citet{mar13}, and FSPS uses the MIST isochrones with empirical TP-AGB spectra from \citet{lan02}.

At longer wavelengths ($\lambda_{\rm rest} > 4~{\rm \mu m}$), as shown in {\color{blue} \textbf{Figures \ref{fig:offset_nE} and \ref{fig:offset_age}}}, the three SPS models other than E-MILES overpredict the observed SEDs by up to $\sim50\%$, while E-MILES exhibits good agreement with the SPHEREx data.
The E-MILES behavior at these wavelengths may be related to its prominent CO absorption features around $4.2-4.5~{\rm \mu m}$, which are modeled in detail compared to the other models \citep{roc15, LeeJH25}.
Although the detailed absorption features of this CO band are difficult to clearly identify in the stacked SEDs due to the lower SPHEREx spectral resolution ($R\sim130$) than the native resolution of E-MILES ($R\sim2000$) at these wavelengths \citep{vaz16}, the inclusion of CO absorption features can account for the SED overpredictions of $\sim20-30\%$ observed in models other than E-MILES (see {\color{blue} \textbf{Appendix B}}).

There are no significant changes in the SED offsets between the metal-poor (median $[Z/Z_{\odot}]=-0.11$) and metal-rich populations (median $[Z/Z_{\odot}]=0.06$), indicating that stellar metallicity has little effect on the overall trends of the SED offsets.
In particular, the modeled SEDs from all SPS models except FSPS are nearly invariant with metallicity.
For FSPS, however, the SED offsets of the metal-rich galaxies are $\sim10-20\%$ larger than those of metal-poor galaxies.
This metallicity dependence may partially contribute to the overprediction of $\sim0.15-0.2~{\rm mag}$ at the wavelength bins of $\lambda_{\rm rest} > 2.42~{\rm \mu m}$ noted in {\color{blue} \textbf{Figure \ref{fig:offset_nE}}}.
Overall, these results suggest that differences in the treatment of NIR spectral features among SPS models are the main drivers of the observed SED offsets in intermediate-age galaxies, rather than the effect of metallicity.

\subsection{Implications for NIR SEDs from SPS Models}
\label{sec:dis3}


In this study, we consider that the observed SED offset trends are genuinely driven by the intrinsic SED differences among SPS models.
Here we consider that there are little systematic biases in SPHEREx flux calibration, in the sense that the offsets show clear correlations with stellar age and ${\rm W1-W3}$ color, and the SPHEREx fluxes are fairly consistent with WISE W1 measurements.
In addition, differences in the derived stellar ages and metallicities among SPS models do not have significant effect on the model dependence of the SED offsets (see {\color{blue} \textbf{Appendix C}}).

Among the SPS models adopted in this study, the E-MILES model can be regarded as a reliable choice for NIR SED predictions.
E-MILES generally shows the best performance for NIR SED predictions across the NIR wavelengths, except for a slight overprediction at $\lambda_{\rm rest}=2.6-3~{\rm \mu m}$ for intermediate-age stellar populations.

One of the main strengths of E-MILES is the incorporation of empirical stellar spectral libraries in the NIR.
The SSP models in the NIR are calculated based on 180 spectra of cool stars from the IRTF library \citep{cus05, ray09}, while other SPS models mainly rely on the theoretical BaSeL library \citep{wes02}. 
Although the empirical libraries have limited coverage in stellar parameters, such as effective temperature and metallicity, the inclusion of IRTF stars is beneficial in reflecting real stellar spectra compared to theoretical frameworks. 

Furthermore, E-MILES provides SSP models up to $5~{\rm \mu m}$ with relatively high spectral resolution ($R \sim 2000$), based on medium-resolution NIR spectroscopy from SpeX \citep{ray03} and supplementary theoretical prediction from the Phoenix library \citep{all12}.
This allows E-MILES to include detailed prescriptions of CO absorption features ($4.2-4.5~{\rm \mu m}$), leading to a better match with observed NIR colors of early-type galaxies than the other models \citep{roc15, roc16}.
This can also explain the good agreement with SPHEREx at $\lambda_{\rm rest}=3.82-5~{\rm \mu m}$, as shown in {\color{blue} \textbf{Figures \ref{fig:offset_nE} and \ref{fig:offset_age}}}.

However, E-MILES tends to overpredict the SEDs by $\sim 0.1-0.2~{\rm mag}$ at $\lambda_{\rm rest}=2.6-3~{\rm \mu m}$ for galaxies dominated by intermediate-age stellar populations.
This overprediction range is consistent with the typical ages of TP-AGB stars ($\sim0.2-2~{\rm Gyr}$), suggestive of the main drivers of the NIR SED discrepancies.
As TP-AGB stars can dominate NIR fluxes in the SEDs \citep{mar98, mar05, ko13, noe13}, this overprediction may be attributed to possible uncertainties in the treatment of TP-AGB stars at wavelengths $\gtrsim2.4~{\rm \mu m}$.
There may be additional uncertainties in modeling carbon stars or oxygen-rich stars, which have intermediate age and low temperature ($T_{e} \sim 2400-4000~{\rm K}$).
These stars exhibit strong absorption features around $\sim3~{\rm \mu m}$ in their spectra, originated from molecules such as CO or ${\rm H_{2}O}$ \citep{tsu97, ari02, ari09}, which may also contribute to the SED overprediction in E-MILES.
The IRTF library has observational limitations to cover these stars, including 32 AGB stars, 5 carbon stars, and no oxygen-rich stars \citep{roc15, vil17}, which may not be sufficient to capture the diverse NIR emissions from intermediate-age stellar populations.

The other SPS models primarily incorporate the lower-resolution BaSeL library for their NIR SEDs \citep{bru03, con09, pla19}, supplemented with additional treatments for TP-AGBs, carbon stars, and extremely hot stars ($T_{e} \gtrsim 50000~{\rm K}$).
Although this theoretical library covers a wide range of stellar temperatures and metallicities, it has limitations in reproducing realistic stellar continua and complex spectral features of cool giant stars, particularly in the NIR wavelengths.
This can lead to the systematic overpredictions of the modeled SEDs at $\lambda_{\rm rest} \gtrsim 2.4~{\rm \mu m}$ across all age ranges, in contrast to the trend of E-MILES for old stellar populations.
A similar discrepancy was also noticed in \citet{LeeJH25}, which found substantial differences in the NIR ${\rm [3.6]-[4.5]}$ color between E-MILES and other models (see their Figure 7).

SPHEREx observations provide a valuable opportunity for a systematic evaluation of the NIR SEDs predicted by SPS models using a large sample of galaxies, such as post-starburst galaxies, which are expected to exhibit similar properties to non-ELGs.
We can further compare the modeled NIR SEDs with SPHEREx observations, using star clusters (e.g., young star clusters or globular clusters), which are closer to simple stellar populations than the sample galaxies in this study.
Taking advantage of its wide sky coverage and dense NIR wavelength sampling, SPHEREx can also offer empirical stellar SEDs for a larger sample of cool stars \citep{bro26, rus26}, which provide valuable frameworks for SED modeling of galaxies in future studies.
Specifically, SPHEREx Bands 5 and 6 cover the CO absorption features in galaxies, which are not well reproduced in the SPS models other than E-MILES.
Even in E-MILES, these CO features are not based on the observed stellar spectra but on the theoretical Phoenix library \citep{all12}, due to the spectral overlap with telluric absorption features \citep{roc15}.

Furthermore, SPHEREx data will enable full SED fitting of galaxies, including both stellar and non-stellar components (e.g., with CIGALE; \citeauthor{boq19} \citeyear{boq19}).
The full SED fitting spanning the optical and NIR wavelength ranges would be essential for accurately constraining the SFHs of galaxies, as the optical and NIR observations probe different stellar populations within galaxies.
This capability can be a strong opportunity for advancing stellar population studies through synergy with the future optical surveys \citep{jes26}, such as the Legacy Survey of Space and Time (LSST; \citeauthor{ive19} \citeyear{ive19}) and 7-Dimensional Sky Survey (7DS; \citeauthor{im24} \citeyear{im24}, see also \citeauthor{lim25} \citeyear{lim25}).
Combining these optical surveys with SPHEREx would be beneficial to explore elusive stellar components over the broad wavelength range, particularly TP-AGB stars, cool giants with molecular absorption features, or very young stars with PAH emission, which are difficult to distinguish with optical broadband data alone.
In addition, in a cosmological context, this approach can be broadly extended to constrain SPS models for galaxies over a wide range of redshifts.

\section{Summary}
\label{sec:summary}

This study investigates the model dependence of NIR SEDs in comparison with SPHEREx observational data.
SPHEREx provides all-sky spectrophotometric data with wide wavelength coverage from $0.75~{\rm \mu m}$ to $5~{\rm \mu m}$, which is beneficial in this comparative analysis.
In this study, we select 3,889 compact galaxies, including 2,726 non-ELGs and 1,163 ELGs, with SDSS optical spectra and corresponding SPHEREx data.
With the sample galaxies, we derive modeled SEDs using full-spectrum fitting with four NIR-covering SPS models, E-MILES, BC03, CB19, and FSPS.
We then quantify the relative differences between the modeled SEDs and SPHEREx observations by calculating the weighted offsets ($\mu_{\Delta}$) depending on galaxy type (non-ELGs and ELGs), SPS model, and SPHEREx band.
We summarize the main findings from this study as follows.

\begin{enumerate}
\item
For both non-ELGs and ELGs, the modeled SEDs are relatively in good agreement with SPHEREx at $\lambda_{\rm rest}=0.75-2.42~{\rm \mu m}$.
At $\lambda_{\rm rest}=2.42-5~{\rm \mu m}$, all SPS models tend to overpredict the NIR SEDs of non-ELGs by $\sim0.1-0.3~{\rm mag}$.
In contrast, the models except for CB19 exhibit substantial underpredictions of the SEDs relative to SPHEREx.
\item
For ELGs, the SED offsets at $\lambda_{\rm rest}=2.42-5~{\rm \mu m}$ are strongly affected by dust emission, which is not included in the modeled SEDs from SPS models.
The clear correlations between $\mu_{\Delta}$ and WISE ${\rm W1-W3}$ colors demonstrate the contribution of non-stellar emission at these wavelengths.
Therefore, the SED offsets of ELGs at $\lambda_{\rm rest}=2.42-5~{\rm \mu m}$ arise from the combined effects of model dependence and non-stellar contribution.
For this reason, we consider non-ELGs as our main analysis sample to explore the model dependence of NIR SED predictions.
\item
For non-ELGs, E-MILES yields the best agreement with SPHEREx among the SPS models.
Although E-MILES overpredicts the SEDs at $\lambda_{\rm rest}\sim2.6-3~{\rm \mu m}$ by $\sim0.1~{\rm mag}$ in median, this offset decreases at $\lambda_{\rm rest}=3.82-5~{\rm \mu m}$.
In contrast, the other models consistently show the overprediction by $0.2-0.3~{\rm mag}$ across $\lambda_{\rm rest}=2.42-5~{\rm \mu m}$.
\item
The SED overprediction of E-MILES is mainly observed in intermediate-age stellar populations.
This may reflect uncertainties in the empirical stellar library (IRTF) adopted in E-MILES, likely due to the limited coverage of intermediate-age stars (e.g., TP-AGB stars, carbon stars, or oxygen-rich stars).
Compared to the other SPS models, the improved performance of E-MILES at $\lambda_{\rm rest}=3.82-5~{\rm \mu m}$ can be attributed to its high-resolution treatment of the CO absorption features at these wavelengths.
\item
The other three SPS models (BC03, CB19, and FSPS) have limitations in their NIR-covering spectral library (BaSeL), which may not be sufficient to reproduce the observed NIR SEDs from cool giant stars.
In particular, the SED overestimations at $\lambda_{\rm rest}=3.82-5~{\rm \mu m}$ may partially result from their lower-resolution NIR spectral libraries, which do not incorporate the CO absorption features, in contrast to E-MILES.
\end{enumerate}

\begin{acknowledgements}
We thank the anonymous referee for numerous constructive comments that have significantly improved the manuscript.
This work was supported by the Basic Science Research Program through the National Research Foundation of Korea (NRF) funded by the Ministry of Education (No. RS-2024-00452816), NRF grant funded by the Korean government (MSIT) (Nos. RS-2024-00347548 and RS-2025-16066624), and the Yonsei University Research Fund of 2025 (2025-22-0402).
K.O. acknowledges support from the National Research Foundation of Korea (NRF) grant funded by the Korea government (MSIT) (RS-2025-00553982).
Bomee Lee is supported by the National Research Foundation of Korea (NRF) grant funded by the Korea government(MSIT), 2022R1C1C1008695.
Y. K. was supported by the National Research Foundation of Korea (NRF) grant funded by the Korean government (MSIT) (No. RS-2026-25476464).
Part of the research described in this paper was carried out at the Jet Propulsion Laboratory, California Institute of Technology, under a contract with the National Aeronautics and Space Administration (80NM0018D0004).

Funding for the Sloan Digital Sky Survey V has been provided by the Alfred P. Sloan Foundation, the Heising-Simons Foundation, the National Science Foundation, and the Participating Institutions. SDSS acknowledges support and resources from the Center for High-Performance Computing at the University of Utah. SDSS telescopes are located at Apache Point Observatory, funded by the Astrophysical Research Consortium and operated by New Mexico State University, and at Las Campanas Observatory, operated by the Carnegie Institution for Science. The SDSS website is \url{www.sdss.org}.

SDSS is managed by the Astrophysical Research Consortium for the Participating Institutions of the SDSS Collaboration, including the Carnegie Institution for Science, Chilean National Time Allocation Committee (CNTAC) ratified researchers, Caltech, the Gotham Participation Group, Harvard University, Heidelberg University, The Flatiron Institute, The Johns Hopkins University, L'Ecole polytechnique f\'{e}d\'{e}rale de Lausanne (EPFL), Leibniz-Institut f\"{u}r Astrophysik Potsdam (AIP), Max-Planck-Institut f\"{u}r Astronomie (MPIA Heidelberg), Max-Planck-Institut f\"{u}r Extraterrestrische Physik (MPE), Nanjing University, National Astronomical Observatories of China (NAOC), New Mexico State University, The Ohio State University, Pennsylvania State University, Smithsonian Astrophysical Observatory, Space Telescope Science Institute (STScI), the Stellar Astrophysics Participation Group, Universidad Nacional Aut\'{o}noma de M\'{e}xico, University of Arizona, University of Colorado Boulder, University of Illinois at Urbana-Champaign, University of Toronto, University of Utah, University of Virginia, Yale University, and Yunnan University.

This research has made use of the NASA/IPAC Infrared Science Archive, which is funded by the National Aeronautics and Space Administration and operated by the California Institute of Technology.
This publication makes use of data products from the Spectro-Photometer for the History of the Universe, Epoch of Reionization and Ices Explorer (SPHEREx), which is a joint project of the Jet Propulsion Laboratory and the California Institute of Technology, and is funded by the National Aeronautics and Space Administration (doi:\dataset[10.26131/IRSA652]{https://doi.org/10.26131/IRSA652}).
This publication makes use of data products from the Wide-field Infrared Survey Explorer, which is a joint project of the University of California, Los Angeles, and the Jet Propulsion Laboratory/California Institute of Technology, funded by the National Aeronautics and Space Administration (doi:\dataset[10.26131/IRSA142]{https://doi.org/10.26131/IRSA142}).

\end{acknowledgements}

\software{Numpy \citep{har20}, Matplotlib \citep{hun07}, Scipy \citep{vir20}, Astropy \citep{ast13, ast18, ast22}, pPXF \citep{cap23}}

\facility{SDSS, IRSA, SPHEREx, WISE}

\appendix

\section{Stacked NIR SEDs of non-ELGs with Old Stellar Populations}
\label{app:A}

\restartappendixnumbering

In this Appendix, we present comparisons of the stacked SEDs for old non-ELGs as a supplement to the intermediate-age results shown in {\color{blue} \textbf{Section \ref{sec:dis1}}}.
Although the SED offsets of old stellar populations are weaker than those of intermediate-age populations, examining their trends would be important for understanding the overall trends of non-ELGs because approximately $80\%$ of the non-ELGs have stellar ages of ${\rm log~Age~(yr) \geq 9.5}$.

\begin{figure}
\centering
\includegraphics[width=1.0\textwidth]{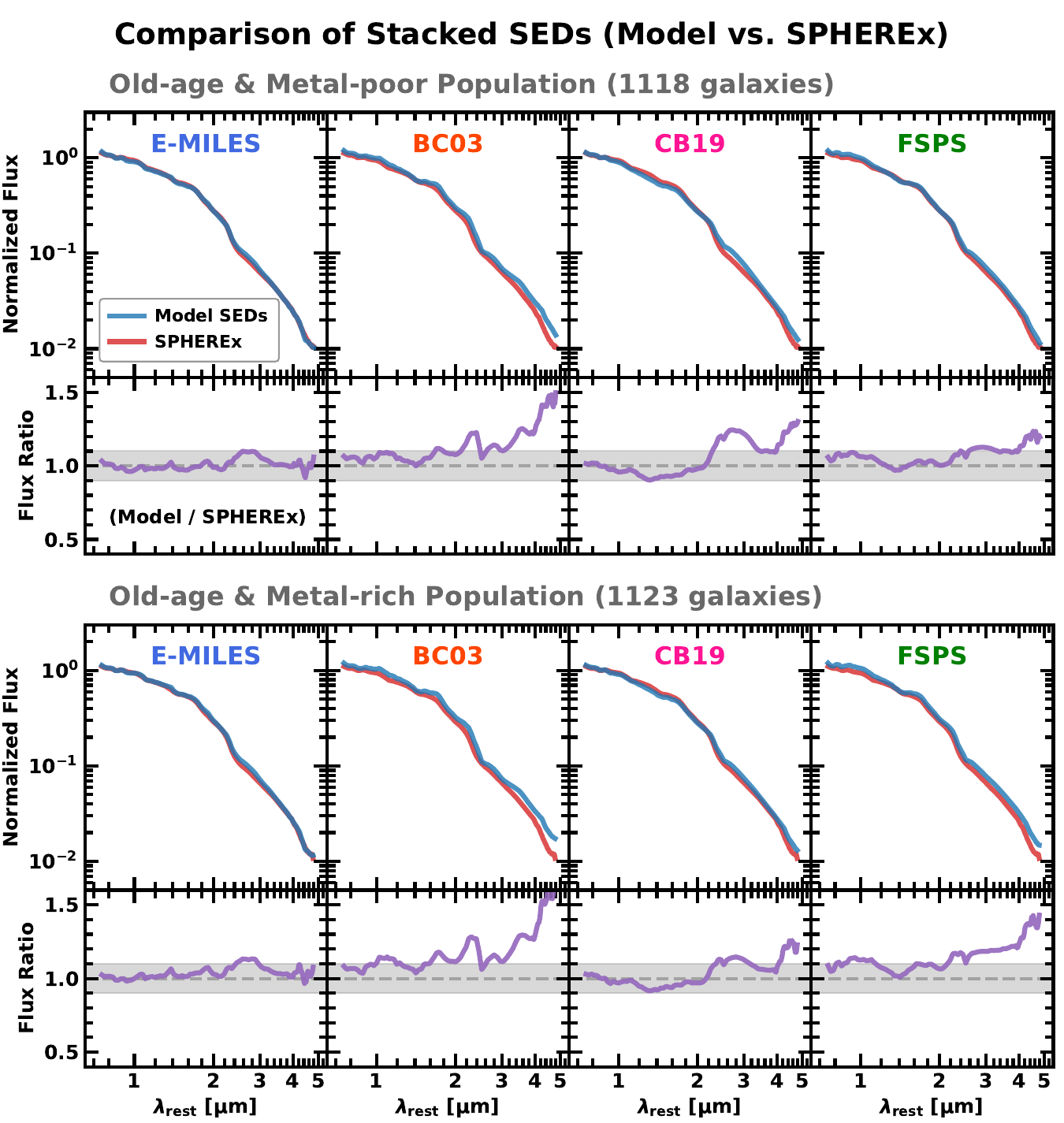}
\caption{
Same as {\color{blue} \textbf{Figure \ref{fig:stack}}}, but for old-age stellar populations.
\label{figA:stack_old}}
\end{figure}

{\color{blue} \textbf{Figure \ref{figA:stack_old}}} compares the stacked SEDs from the SPS models and SPHEREx for non-ELGs with ${\rm log~Age~(yr) \geq 9.5}$, divided into two metallicity bins based on the median metallicity from E-MILES ($[Z/Z_{\odot}]=-0.12$): metal-poor (median $[Z/Z_{\odot}]=-0.22$) and metal-rich (median $[Z/Z_{\odot}]=-0.03$) populations, as done in {\color{blue} \textbf{Figure \ref{fig:stack}}}.
The general trends of SED overprediction observed in intermediate-age galaxies also appear over similar spectral ranges.
For E-MILES, the SED overpredictions peak at $\lambda_{\rm rest} \sim 2.6-3~{\rm \mu m}$, consistent with the intermediate-age galaxies, but their amplitudes are mitigated to $\sim10\%$, about half of those found in the intermediate-age population.
The CB19 model shows SED overpredictions of $\sim10-20\%$, which are substantially smaller than those for intermediate-age galaxies ($\sim40\%$).
In contrast, the BC03 and FSPS models show little difference between the old-age and intermediate-age populations, as presented in {\color{blue} \textbf{Figure \ref{fig:offset_age}}}.
As in the intermediate-age non-ELGs, stellar metallicity has little effect on the overprediction trends.

\section{Comparisons of Model SSPs at NIR}
\label{app:B}

\restartappendixnumbering

To explore the intrinsic differences in NIR SEDs, we compare the original SSP templates from the SPS models.
While the stacked NIR SEDs presented in {\color{blue} \textbf{Figure \ref{fig:stack}}} represent composite stellar populations constructed by combining SSP templates with different ages and metallicities, the individual SSP templates provide the fundamental SEDs at fixed stellar ages and metallicities, which serve as ingredients of the composite SEDs used in our analysis.

\begin{figure}
\centering
\includegraphics[width=1.0\textwidth]{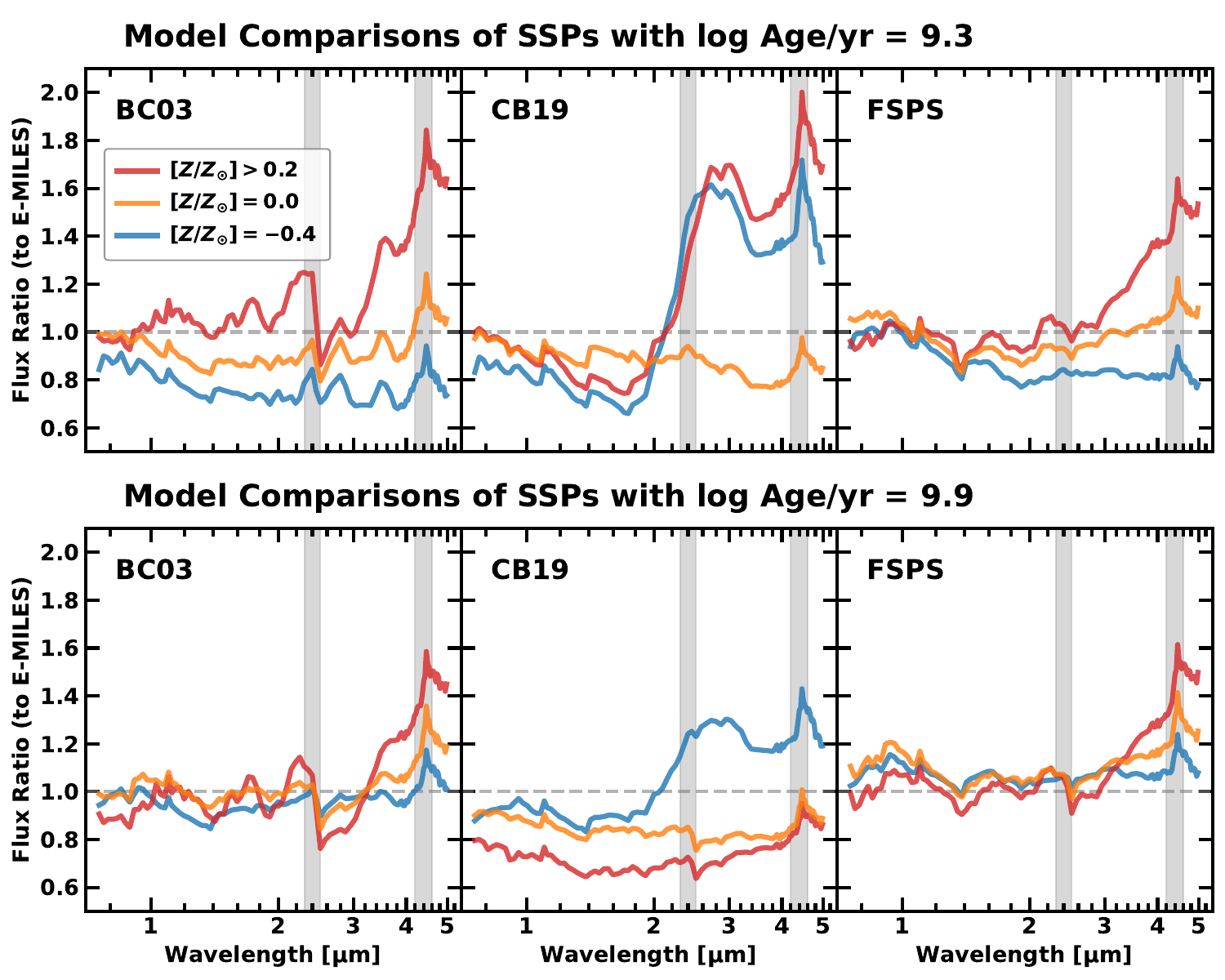}
\caption{
Comparisons of simple stellar populations (SSPs) among the four SPS models adopted in this study.
The figure plots the flux ratios of the SSP spectral templates from BC03 (left column), CB19 (middle column), and FSPS (right column) relative to those from E-MILES.
The upper three panels correspond to ${\rm log~Age/yr=9.3}$, and the lower panels correspond to ${\rm log~Age/yr=9.9}$.
In each panel, the flux ratios are displayed for three different metallicity bins: $[Z/Z_{\odot}]>0.2$ (red), $[Z/Z_{\odot}]=0.0$ (orange), and $[Z/Z_{\odot}]=-0.4$ (blue).
The gray-shaded regions indicate the CO first overtone band ($\sim2.3-2.5~{\rm \mu m}$ and the fundamental band ($\sim4.2-4.5~{\rm \mu m}$).
\label{figA:ssp}}
\end{figure}

{\color{blue} Figure \ref{figA:ssp}} displays the SSP comparisons of the three SPS models (BC03, CB19, and FSPS) with those of E-MILES, with the spectral resolution of each SSP template downgraded to the SPHEREx resolution using the 102 SPHEREx wavelength channels.
The two age bins (${\rm log~Age/yr=9.3}$ and ${\rm log~Age/yr=9.9}$) correspond to the median stellar ages of non-ELGs in the intermediate-age and old-age bins, respectively.
In each panel, the flux ratios of the SSP templates relative to E-MILES are plotted with three metallicity bins: $[Z/Z_{\odot}]=-0.4$, $[Z/Z_{\odot}]=0.0$, and the highest metallicity bin $[Z/Z_{\odot}]>0.2$.
Note that the highest metallicity bins are slightly different among the SPS models: $[Z/Z_{\odot}]=0.22$ for E-MILES, $[Z/Z_{\odot}]=0.3$ for CB19, and $[Z/Z_{\odot}]=0.4$ for BC03 and FSPS.

As expected, the SSPs from the three models show relatively small differences with those from those of E-MILES at $\lambda_{\rm rest}<2.4~{\rm \mu m}$, with most templates exhibiting slight flux underpredictions compared to E-MILES.
At $\lambda_{\rm rest}>2.4~{\rm \mu m}$, where the SED overpredictions and model dependence primarily appear in the composite SEDs, the SSPs tend to show substantially larger offsets from the E-MILES templates, depending on the stellar age and metallicity bins.
The BC03 and FSPS models yield increasingly larger offsets from E-MILES toward older stellar ages and higher metallicities.
For CB19, the dependence on metallicity seems more complex, showing larger flux overestimations than E-MILES in the low-metallicity bins at both ages, as well as in the highest metallicity bin at the intermediate age.
This may reflect the combined effect of the adopted stellar evolutionary tracks and the treatment of cool evolved stars in CB19, including the PARSEC isochrones and dusty stellar envelope models for the TP-AGB phase \citep{gon10}, which may lead to significantly different trends of metal-poor cool stars compared to other models.
These SSP-level behaviors combine to produce the SED overpredictions observed in the composite SEDs shown in {\color{blue} \textbf{Figures \ref{fig:offset_nE} and \ref{fig:offset_age}}}.

In terms of the CO absorption features, the first-overtone CO band ($\sim2.3-2.5~{\rm \mu m}$) is difficult to identify, while the fundamental CO band ($\sim4.2-4.5~{\rm \mu m}$) is clearly visible in the SSP comparisons.
These features appear as prominent peaks in the SSP flux ratios relative to E-MILES because the three SPS models do not incorporate this CO absorption band.
Comparing the flux ratios around the CO band, the three models seem to overestimate the fluxes by approximately $20\%$ compared to E-MILES.
This implies that the differences in the SED offsets between E-MILES and the other SPS models can be effectively explained by the inclusion of the CO absorption band at $\sim 4.2-4.5~{\rm \mu m}$ in E-MILES.

\section{Coefficients of Determination for the SED Offsets}
\label{app:C}

\restartappendixnumbering

As discussed in {\color{blue} \textbf{Section \ref{sec:res1}}}, the overall distributions of stellar age and metallicity derived using pPXF are similar among the adopted SPS models.
This indicates that the inferred stellar population properties do not exhibit significant systematic biases depending on the choice of SPS model.
However, the derived ages and metallicities are not completely identical across the SPS models and may also be affected by the age--metallicity degeneracy, which potentially introduces additional dependencies of the SED offsets on these stellar population parameters.

To quantify the degree to which the differences in the inferred age and metallicity contribute to the model dependence of the SED offsets shown in {\color{blue} \textbf{Figures \ref{fig:offset_age} and \ref{fig:stack}}}, we perform a simple linear regression analysis.
We construct a linear model in which the difference in the SED offsets ($\mu_{\Delta}$) between two SPS models is treated as the dependent variable, and the corresponding differences in stellar age and metallicity are considered as explanatory variables:
\begin{equation}
    \mu_{\Delta}(A)-\mu_{\Delta}(B)=C_{0}+C_{Age}\times \Delta {\rm log(Age/yr)}+C_{Z}\times \Delta [Z/Z_{\odot}]
\end{equation}
where $\mu_{\Delta}(A)-\mu_{\Delta}(B)$ represents the difference in the SED offsets predicted by SPS models $A$ and $B$.
$\Delta {\rm log(Age/yr)}$ and $\Delta [Z/Z_{\odot}]$ represent the differences in the stellar age and metallicity derived from the two different models, respectively.
$C_{X}$ denotes the regression coefficients.
By performing the multi-linear regression independently at each wavelength, we determine the coefficient of determination ($R^{2}$) to quantify the fraction of the variance in the SED offset differences that can be explained by differences in the computed stellar age and metallicity among the SPS models.
It is important to note that the $R^{2}$ values assess the strength of the correlations between the variables and do not measure the absolute magnitude of the SED offsets.

\begin{figure}
\centering
\includegraphics[width=1.0\textwidth]{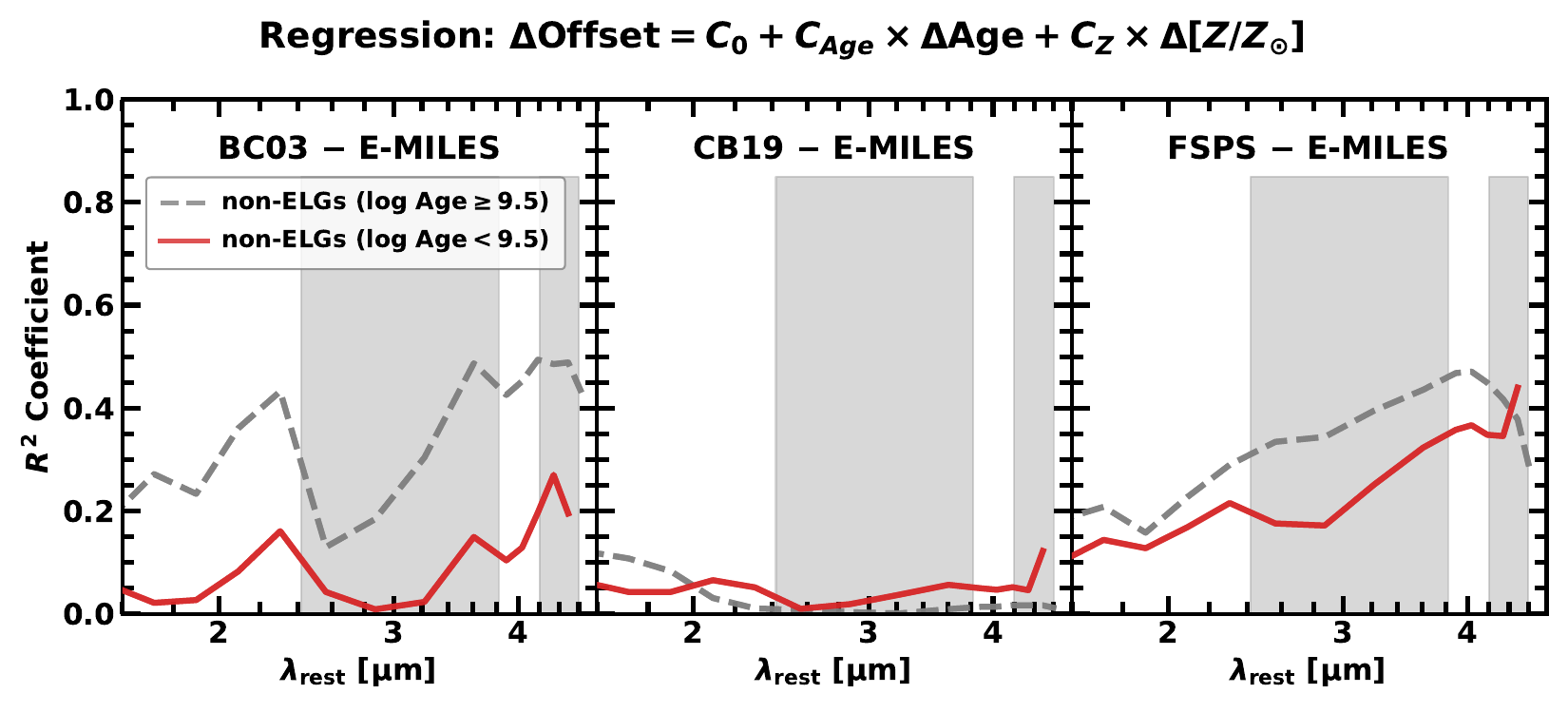}
\caption{
Here we perform a regression analysis to explore how differences in stellar age and metallicity derived from different SPS models contribute to the model-dependent SED offsets.
To quantify the explanatory power of the age and metallicity differences, we use the coefficient of determination ($R^{2}$) from the regression model noted at the top.
In this figure, each panel shows the distributions of the $R^{2}$ values as a function of rest-frame wavelength.
The left, middle, and right panels compare BC03, CB19, and FSPS with E-MILES, respectively.
Within each panel, the results are shown separately for intermediate-age non-ELGs (${\rm log~Age/yr < 9.5}$; red solid line) and old-age non-ELGs (${\rm log~Age/yr \geq 9.5}$; gray dashed line).
The shaded regions represent the rest-frame wavelength ranges of $\lambda_{\rm rest}=2.42-3.82~{\rm \mu m}$ and $\lambda_{\rm rest}=4.2-4.6~{\rm \mu m}$, where the modeled SED offsets show substantial differences relative to E-MILES.
The $R^{2}$ distributions suggest that the differences in the SED offsets among the SPS models are primarily driven by intrinsic differences in the modeled SEDs, rather than by variations in the stellar age and metallicity derived from the different SPS models.
\label{figA:R2}}
\end{figure}

{\color{blue} \textbf{Figure \ref{figA:R2}}} shows the resulting $R^{2}$ coefficients as a function of wavelength.
In this analysis, we adopt the E-MILES model as the reference and perform the regression separately for each of the other SPS models at $\lambda_{\rm rest}>1.64~{\rm \mu m}$, where model-dependent SED offsets begin to emerge.
Overall, the $R^{2}$ values remain below 0.4, indicating that differences in stellar age and metallicity can explain less than $40\%$ of the variance in the SED offset differences over these wavelengths.

Intermediate-age non-ELGs tend to exhibit lower $R^{2}$ values than old galaxies at these wavelengths for all three models.
In particular, around the E-MILES SED overprediction peaks ($\lambda_{\rm rest}\sim2.6-3~{\rm mu m}$), the other three SPS models show significantly low $R^{2}$ values for intermediate-age non-ELGs, reaching nearly zero for BC03 and CB19 and only $\sim0.2$ for FSPS.
Similarly, within the CO absorption band ($\sim 4.2-4.5~{\rm \mu m}$), the $R^{2}$ values remain low at approximately 0.2, 0.1, and 0.4 for BC03, CB19, and FSPS, respectively.
These results demonstrate that the SED offsets, particularly those of intermediate-age populations at $\lambda_{\rm rest}>2.4~{\rm \mu m}$, mostly arise from intrinsic differences among the SPS model SEDs, rather than from differences in the derived stellar ages and metallicities.

\end{document}